\documentclass[twocolumn]{autart}    

\usepackage{graphicx}          
                               
\usepackage{cite}
\usepackage{epsfig}
\usepackage{float}
\usepackage{subfigure}
\usepackage{times}
\usepackage{amsmath}
\usepackage{amssymb}
\usepackage{multirow}
\usepackage{stfloats}
\usepackage{xcolor}
\usepackage{cite}
\usepackage{epsfig}
\usepackage{float}
\usepackage{subfigure}
\usepackage{times}
\usepackage{amsmath}
\usepackage{url}
\newtheorem{theorem}{Theorem}
\newtheorem{lemma}{Lemma}
\newtheorem{remark}{Remark}
\newtheorem{definition}{Definition}
\newtheorem{assumption}{Assumption}
\usepackage{amssymb}
\usepackage{multirow}
\usepackage{hhline}
\usepackage{arydshln}
\usepackage{autobreak}

\usepackage{amsfonts}
\usepackage{algorithmic}
\usepackage{algorithm}
\usepackage{array}
\usepackage{textcomp}
\usepackage{verbatim}
\usepackage{placeins}
\usepackage{bm}
\usepackage{makecell}
\allowdisplaybreaks

\begin{document}

\begin{frontmatter}

\title{Distributed Frequency Control for Multi-Area Power Systems Considering Transient Frequency Safety \thanksref{footnoteinfo}} 

\thanks[footnoteinfo]{This paper was not presented at any IFAC 
meeting. Corresponding author T. Liu. The work was supported by the National Natural Science Foundation of China through Project No. 62173287 and the Research Grants Council of the Hong Kong Special Administrative Region under the Early Career Scheme through Project No. 27206021.}

\author[hku,SIRI]{Xiemin Mo}\ead{xieminmo@eee.hku.hk},          
\author[hku,SIRI]{Tao Liu}\ead{taoliu@eee.hku.hk},               

\address[hku]{Department of Electrical and Electronic Engineering, University of Hong Kong, Hong Kong}  
\address[SIRI]{Shenzhen Institute of Research and Innovation, The University of Hong Kong, Shenzhen, China}             

\begin{keyword}                           
Distributed optimization/control, frequency regulation, transient frequency safety.           
\end{keyword}                             

\begin{abstract}                          
High penetration of renewable energy sources intensifies frequency fluctuations in multi-area power systems, challenging both stability and operational safety. This paper proposes a novel distributed frequency control method that ensures transient frequency safety and enforces generation capacity constraints, while achieving steady-state frequency restoration and optimal economic operation. The method integrates a feedback optimization (FO)-based controller and a safety corrector. The FO-based controller generates reference setpoints by solving an optimization problem, driving the system to the steady state corresponding to the optimal solution of this problem. The safety corrector then modifies these references using control barrier functions to maintain frequencies within prescribed safe bounds during transients while respecting capacity constraints. The proposed method combines low computational burden with improved regulation performance and enhanced practical applicability. Theoretical analysis establishes optimality, asymptotic stability, and transient frequency safety for the closed-loop system. Simulation studies show that, compared with conventional FO-based schemes, the method consistently enforces frequency safety and capacity limits, achieves smaller frequency deviations and faster recovery, thereby demonstrating its practical effectiveness and advantages.
\end{abstract}

\end{frontmatter}
\section{Introduction}\label{introduction}
Large-scale power systems are generally composed of multiple interconnected subsystems, with each subsystem corresponding to a regional grid or administrative zone \cite{7544628,9495219}. These subsystems are referred to as control areas. To ensure the safe and stable operation of power systems, frequency regulation aims to keep the system frequency and net tie-line powers at their nominal values \cite{kundur1994}. In conventional power systems dominated by synchronous generators (SGs), automatic generation control (AGC) has been widely used to achieve these two goals over the past several decades \cite{Wood2014}. However, the increasing integration of renewable energy sources has introduced greater variability and reduced system inertia, posing new challenges to frequency regulation.

Due to the significant volatility of renewables, AGC may not be able to respond promptly to rapid and significant fluctuation in net loads (i.e., load minus renewable generation), resulting in a deterioration of the system frequency \cite{DORFLER2017296}. To deal with this issue, various distributed frequency control methods have been proposed to provide fast control actions. These methods can be roughly divided into the reinforcement learning (RL)-based method (e.g., \cite{Yan2020,Zhao2022,10696981}), model predictive control (MPC)-based method (e.g., \cite{Yang2021b,Liu2023}), and feedback optimization (FO)-based method (e.g., \cite{Trip2016,Trip2018,9869334,Li2016,7944568,Ste2017,Yang2021,10319778,zhao2024distributed,wang2020distributed,10488734}). Specifically, the RL-based and MPC-based methods can optimize both transient and steady-state performance but often incur high computational costs from offline training or real-time optimization, limiting their scalability in large-scale systems \cite{Yan2020,10696981,Yang2021}. In contrast, the FO-based method strikes a balance between control performance and computational efficiency. By formulating a steady-state optimization problem and designing a controller with an analytical control update law, this method drives the system to the optimal solution automatically. It requires no offline training, offers low computational complexity, and remains practically applicable, albeit with some sacrifice in transient optimality \cite{10085973}.

Despite these advantages, the existing FO-based methods typically overlook transient frequency safety. As the share of renewables increases, reduced system inertia leads to a lower frequency nadir and a higher rate of change of frequency (RoCoF), making frequency excursions more severe \cite{dorfler2023control}. Moreover, the volatility of net loads can cause the transient frequency outside safe bounds, risking generator tripping, load shedding, or even cascading failures \cite{8450880}. Consequently, incorporating transient frequency safety into frequency control strategies has become an essential challenge in power systems with high penetration of renewables \cite{10005839}. Centralized approaches, e.g., \cite{10005839}, construct neural network-based barrier certificates offline and apply them online for safe frequency regulation. In response, distributed safety-guaranteed frequency controllers have been proposed, e.g., in \cite{ZHANG2019274,ZHANG2021109335,9031310,YUAN2024105753}. Nonetheless, these distributed methods can only keep frequency within the safety region, but cannot restore it to its nominal value.

Another practical requirement is the enforcement of generator capacity constraints. Owing to the physical constraints of generation units, their output capacities are inherently limited. However, many existing studies either neglect these capacity constraints (e.g., \cite{Yan2020,Trip2016,10696981,Liu2023,Trip2018,9869334,Li2016,7944568,10005839,ZHANG2019274,YUAN2024105753}) or only enforce them at the steady state (e.g., \cite{Ste2017,Yang2021,10319778,zhao2024distributed,wang2020distributed}). When the control command exceed the allowable operating range of a device, the actual generation may deviate from the desired value. This mismatch not only undermines the effectiveness of frequency regulation but also introduces potential risks to system reliability and operational safety \cite{7990543}. Therefore, incorporating such constraints into frequency control strategies is essential for improving the practical feasibility of frequency control and the system safety.

To address the aforementioned challenges, this paper proposes a new distributed control method that ensures both transient frequency safety and capacity constraints while steering the system toward the economically optimal steady state. The control strategy consists of two layers, i.e., an FO-based controller and a safety corrector. The FO-based controller regulates the system frequency and net tie-line powers while driving the system to an optimal solution of an optimization problem. Meanwhile, the safety corrector ensures the frequency safety and capacity constraint during transient procedures. Theoretical analysis is provided to establish the optimality, asymptotic stability, and transient frequency safety of the closed-loop system. Our key contributions are as follows.

(1) The proposed method achieves significant improvements in computational efficiency compared to the RL-based and MPC-based methods (e.g., \cite{Yan2020,Zhao2022,10696981,Yang2021b,Liu2023}). Additionally, it also shows better control performance in terms of regulation time and transient frequency deviation than the traditional FO-based method (e.g., \cite{Trip2016,Trip2018,9869334,Li2016,7944568,Ste2017,Yang2021,10319778,zhao2024distributed,wang2020distributed,10488734}).

(2) The proposed method manages to ensure both the steady-state optimality and transient frequency safety, outperforming the existing FO-based methods that typically overlook transient safety considerations. (e.g., \cite{Trip2016,Trip2018,9869334,Li2016,7944568,Ste2017,Yang2021,10319778,zhao2024distributed,wang2020distributed,10488734}).

(3) The proposed method guarantees the capacity constraint both during transients and at the steady state, unlike most prior FO-based works that either ignore this constraint (e.g., \cite{Trip2016,Trip2018,9869334,Li2016,7944568}) or only consider steady-state capacity constraint (e.g., \cite{Ste2017,Yang2021,10319778,zhao2024distributed,wang2020distributed}), thus improving practical applicability and system safety.

The remainder of this paper is structured as follows. Section 2 describes the problem formulation. Section 3 presents the preliminaries. Section 4 introduces the proposed frequency control method. Section 5 conducts the theoretical analysis of the closed-loop system. Section 6 verifies the proposed method through simulations. Finally, Section 7 gives the conclusion.

\textit{Notations:} Let $\mathbb{R}$, $\mathbb{R}_+$, and $\mathbb{R}_{-}$ denote the sets of real numbers, positive real numbers, and negative real numbers, respectively. Let $\mathbb{R}^{n}$ and $\mathbb{R}^{m \times n}$ represent the sets of $n$-dimensional real vectors and $(m \times n)$-dimensional real matrices, respectively. Notation $\text{diag}\{m_1,...,m_k\}$ represents the diagonal matrix with $m_i \in \mathbb{R}$ for $i = 1,...,k$; $\bm{0}$ is the zero vector with appropriate dimensions; $\bm{1}_N$ is the $N$-dimensional vector with all entries being 1. $\|\cdot\|$ denotes the Euclidean norm of a vector. Given a set $\mathcal{A} \subset \mathbb{R}^n$, $\text{cl} \mathcal{A}$ denotes its closure; $\mathcal{A}_1 \times \cdots \times \mathcal{A}_n$ denotes the Cartesian product of sets $\mathcal{A}_1,..., \mathcal{A}_n$. For any vectors $\bm{u} = [u_1,...,u_n] \in \mathbb{R}^n$ and $\bm{x} = [x_1,...,x_n]^{\top} \in \mathbb{R}^n$, $\bm{u} \le \bm{x}$ denotes $u_i \le x_i, \forall i = 1,...,n$. For any matrix $\bm{A} \in \mathbb{R}^{m \times n}$, $[\bm{A}]_i \in \mathbb{R}^{1 \times n}$ denotes the $i$-th row of $\bm{A}$. Given a differentiable function $g: \mathbb{R}^n \rightarrow \mathbb{R}$, $\frac{\partial g(\bm{x})}{\partial \bm{x}} \in \mathbb{R}^{1 \times n}$ denotes its partial derivative. For any vectors $\bm{u}$, $\bm{x}$, and given bounds $\underline{\bm{x}}, \overline{\bm{x}} \in \mathbb{R}^n$ satisfying $\underline{\bm{x}} \le \bm{x} \le \overline{\bm{x}}$, the projection operator $[\bm{u}]_{\bm{x} - \underline{\bm{x}}}^{\overline{\bm{x}} - \bm{x}}$ is defined component-wise as $[\bm{u}]_{\bm{x} - \underline{\bm{x}}}^{\overline{\bm{x}} - \bm{x}} = [[u_1]_{x_1 - \underline{x}_1}^{\overline{x}_1 - x_1},...,[u_n]_{x_n - \underline{x}_n}^{\overline{x}_n - x_n}]^{\top}$, where for each $i = 1,...,n$, $[u_i]_{x_i - \underline{x}_i}^{\overline{x}_i - x_i} = 0$, if $x_i - \underline{x}_i = 0$ and $u_i \le 0$ or $\overline{x}_i - x_i = 0$ and $u_i \ge 0$; otherwise $[u_i]_{x_i - \underline{x}_i}^{\overline{x}_i - x_i} = u_i$. For any vectors $\bm{u}$ and $\bm{x}$, the component-wise maximum (resp. minimum) operator is defined as $\max\{\bm{u}, \bm{x}\} = [\max\{u_1,x_1\},...,\max\{u_n,x_n\}]^{\top}$ (resp. $\min\{\bm{u}, \bm{x}\} = [\min\{u_1,x_1\},...,\min\{u_n,x_n\}]^{\top}$).

\section{Problem Formulation}\label{formulation}
In this section, we present the studied power network model and then introdcue the control targets.
\subsection{Power Network Model}
Consider a transmission network composed of $N$ control areas, indexed by $\mathcal{N} = \{1,...,N\}$. Each area $i \in \mathcal{N}$ aggregates SGs, renewable energy generators (REGs), and loads, and is equipped with a local control center that regulates frequency and net tie-line power. The areas are interconnected through tie-lines indexed by $\mathcal{M} = \{1,...,M\}$, where each tie-line $(i,j) \in \mathcal{M}$ connects areas $i$ and $j$. For convenience, we use $e$ and $(i,j)$ interchangeably to denote a tie-line and assign an arbitrary direction to each.

We adopt the following standard assumptions for the multi-area system, as widely accepted in the literature (e.g., \cite{Wood2014}): (1) Tie-lines are lossless. (2) Voltage magnitudes remain constant within each control area. (3) Reactive power has no impact on voltage phase angles or system frequency.

To analyze the system behavior following a disturbance, we linearize the dynamics around a pre-disturbance equilibrium point, with all variables treated as deviations. Specifically, we adopt the classic linearized swing equation to describe the frequency dynamics of the equivalent aggregated SG for each area \cite{Wood2014}. We use the constant power model for loads \cite{Wood2014} and regard REGs as negative loads, which is widely adopted in the literature, e.g., in \cite{Yan2020, Trip2016, Trip2018}. 

Let $\theta_i$, $w_i$, $P_{g,i}$ denote the power angle, frequency, and mechanical power input of area $i \in \mathcal{N}$, respectively. Define the reduced power angle $\alpha_i = \theta_i - \theta_N$, $\forall i \in \mathcal{N}$ with $\alpha_N = 0$. The system dynamics are then described by
\begin{subequations}
  \begin{align}
     \dot{\alpha}_{i} =& w_i - w_N, i \in \mathcal{N} \setminus \{N\}  \label{eq:PowerSysOG_Dynamicsa}\\
     M_{i}\dot{w}_{i} =& -D_{i}w_{i} + P_{g,i} - d_i - \sum_{j \in \mathcal{N}_i }B_{ij}(\alpha_i - \alpha_j),  i \in \mathcal{N}. \label{eq:PowerSysOG_Dynamicsb}
  \end{align} \label{eq:PowerSysOG_Dynamics}%
\end{subequations}
Here, for each area $i \in \mathcal{N}$, $M_i \in \mathbb{R}_+$ is the aggregated inertial of the equivalent generating unit, while $D_i \in \mathbb{R}_+$ is the aggregated damping. The net load $d_i$ is defined as the difference between load consumption and power generation of REGs. $\mathcal{N}_i$ is the set of neighbors of area $i$. For each line $(i,j) \in \mathcal{M}$, $B_{ij} \in \mathbb{R}_+$ is the sensitivity of the power flow to angle differences. Please refer to \cite{Wood2014} for more details about these parameters.

\subsection{Control Targets}
Suppose a step change in the net load induces a power imbalance, leading to deviations in system frequency. The goal is to design a distributed frequency controller for each control area $i \in \mathcal{N}$ that achieves the following steady-state objectives
\begin{subequations}
  \begin{align}
    w_i^s &= 0 \label{steady_requirement1} \\
    P_{t,i}^s &= 0, \label{steady_requirement2}
  \end{align} \label{steady_requirement}%
\end{subequations}
where $w_i^s$, $P_{t,i}^s$ are the system frequency and net tie-line power of area $i$ at the steady state, respectively.

As discussed in the Section \ref{introduction}, compared to the traditional SG-dominated systems, the low-inertia systems have lower frequency nadirs and larger RoCoF, making these systems more sensitive to disturbances. Moreover, real-world deployments must consider generation capacity constraints and limited communication capabilities. Thus, to ensure practical and secure frequency regulation, the proposed control framework must meet the following requirements:
\begin{itemize}
  \item \textit{(i) Steady-state economic optimality:} Controllers across all areas should collaboratively minimize the total generation cost at the steady state.
  \item \textit{(ii) Transient frequency safety:} For each area $i \in \mathcal{N}$, let $\underline{w}_i \in \mathbb{R}_{-}$, $\overline{w}_i \in \mathbb{R}_{+}$ be the lower and upper bounds of safe frequency, respectively. If the initial system frequency $w_i(0) \in [\underline{w}_i, \overline{w}_i]$, then the frequency $w_i(t)$ must remain within this interval for all $t > 0$. If $w_i(0) \notin [\underline{w}_i, \overline{w}_i]$, the frequency must enter and remain within the safe region as quickly as possible.
  \item \textit{(iii) Capacity constraints satisfaction:} For each area $i \in \mathcal{N}$, the mechanical power input $P_{g,i}(t)$ must satisfy $P_{g,i}(t) \in [\underline{P}_{g,i}, \overline{P}_{g,i}]$ for all $t \ge 0$.
  \item \textit{(iv) Distributed manner:} Each control area should compute its control signals using only local information and shared information from directly connected neighbors.
\end{itemize} 

\section{Preliminaries}\label{preliminary}
As established in Section \ref{formulation}, the proposed controller must ensure both transient frequency safety and capacity limits. To systematically incorporate these requirements into the control framework, we review three mathematical tools, including projected dynamical systems to handle capacity limits, set invariance to define safe regions, and control barrier functions (CBFs) to enforce transient safety in feedback control.

\subsection{Projected Dynamical System}
In practice, control actions, such as generator power adjustments, are limited by physical capacity constraints. To model these bounds directly in the system dynamics, we adopt the framework of projected dynamical systems.

Let $\bm{x} \in \mathbb{R}^n$ be the system state, constrained within known lower and upper bounds $\underline{\bm{x}}$, $\overline{\bm{x}} \in \mathbb{R}^n$, such that $\underline{\bm{x}} \le \overline{\bm{x}}$. Each component of $\underline{\bm{x}}$ (resp. $\overline{\bm{x}}$) is allowed to take the value $-\infty$ (resp. $+\infty$), if necessary \cite{YI2016259}. Define the admissible state set
\begin{equation*}
  \mathcal{X} = \left\{\bm{x}| \underline{\bm{x}} \le \bm{x} \le \overline{\bm{x}}\right\} \subseteq \mathbb{R}^n,
\end{equation*}
which is closed and convex. Let $G: \mathcal{X} \rightarrow \mathbb{R}^n$ be a Lipschitz continuous vector field. The projected dynamical system associated with $G$ is then given by 
\begin{equation}\label{PDS}
  \dot{\bm{x}} = \left[G(\bm{x})\right]^{\overline{\bm{x}} - \bm{x}}_{\bm{x} - \underline{\bm{x}}},
\end{equation}
where the projection operator $[\cdot]^{\overline{\bm{x}} - \bm{x}}_{\bm{x} - \underline{\bm{x}}}$ ensures that trajectories of $\bm{x}$ remain within $\mathcal{X}$. Unlike systems described by classical ordinary differential equations, system \eqref{PDS} is discontinuous due to the projection, and thus solutions are interpreted in the Carathéodory sense as below.
\begin{definition}[Carathéodory Solution \cite{9075378}]
  A Carathéodory solution to \eqref{PDS} is an absolutely continuous function $\bm{x}: [0, T) \rightarrow \mathcal{X}$ for some $T > 0$, with initial condition $\bm{x}(0) = \bm{x_0} \in \mathcal{X}$, such that \eqref{PDS} holds for almost all $t \in [0, T)$.
\end{definition}

We now present two fundamental results that ensure well-posedness and convergence of projected dynamical systems.
\begin{lemma}[Existence and Uniqueness \cite{Cojocaru2004}]\label{PDS_existence}
  Consider system \eqref{PDS}, where $\mathcal{X}$ is closed and convex, and $G: \mathcal{X} \rightarrow \mathbb{R}^n$ is Lipschitz continuous. Then, for every initial condition $\bm{x_0} \in \mathcal{X}$, there exists a unique Carathéodory solution $\bm{x}: [0, T) \rightarrow \mathcal{X}$ for some $T > 0$.
\end{lemma}

\begin{definition}[Largest Weakly Invariant Set \cite{Hauswirth2021}]
  Consider system \eqref{PDS} with a state set $\mathcal{X}_1 \subset \mathcal{X}$. The set $\mathcal{X}_1$ is called weakly invariant if, for every initial condition $\bm{x}_0 \in \mathcal{X}_1$, there exists a complete solution $\bm{x}(t)$ such that $\bm{x}(t) \in \mathcal{X}_1$ for all $t \ge 0$. The union of all weakly invariant sets is referred to as the largest weakly invariant set.
\end{definition}

\begin{lemma}[Invariance Principle \cite{9075378}]\label{PDS_invariance}
  Let $W: \mathbb{R}^n \rightarrow \mathbb{R}$ be a continuously differentiable function, and define the compact sublevel set $\mathcal{F}_l = \{\bm{x} \in \mathcal{X}| W(\bm{x}) \le l\}$ with $l > 0$. If $\dot{W}(\bm{x}) \le 0$ along trajectories of \eqref{PDS}, then any solution starting in $\mathcal{F}_l$ remains in $\mathcal{F}_l$ and asymptotically converges to the largest weakly invariant set within $\text{cl}\{\bm{x} \in \mathcal{X} | \dot{W}(\bm{x}) = 0\}$.
\end{lemma}

These results ensure that once input constraints are enforced via projection, the system remains well-posed and stable under suitable Lyapunov-like conditions.

In addition, the projection operator in \eqref{PDS} exhibits the following structural property.
\begin{lemma}[
\cite{9075378}] \label{property_projection}
  Consider system \eqref{PDS}. There exists a unique vector $\bm{e} \in \mathcal{N}_{\mathcal{X}}(\bm{x})$ such that $\left[G(\bm{x})\right]^{\overline{\bm{x}} - \bm{x}}_{\bm{x} - \underline{\bm{x}}} = G(\bm{x}) - \bm{e}$, where $\mathcal{N}_{\mathcal{X}}(\bm{x}) = \{\bm{e}| \bm{e}^{\top}(\bm{x}' - \bm{x}) \le 0, \forall \bm{x}' \in \mathcal{X}\}$ is the normal cone to set $\mathcal{X}$ at $\bm{x}$ \cite{Cojocaru2004}.
\end{lemma}

Based on this, we further establish the following properties of projected dynamical systems.
\begin{lemma}
\label{lemma_projection}
  Consider the following system
  \begin{subequations} \label{lemma_projection_dynamicsys}%
    \begin{align}
      \dot{\bm{s}} &= [f_1(\bm{s},\bm{y})]_{\bm{s} - \underline{\bm{s}}}^{\overline{\bm{s}} - \bm{s}}  \label{lemma_projection_dynamicsys1} \\
      \dot{\bm{y}} &= [f_2(\bm{s},\bm{y})]_{\bm{y} - \underline{\bm{y}}}^{\overline{\bm{y}} - \bm{y}}, \label{lemma_projection_dynamicsys2}
    \end{align} 
  \end{subequations}
  where $\bm{s} \in \mathcal{S} \subseteq \mathbb{R}^{p}$, $\bm{y} \in \mathcal{Y} \subseteq \mathbb{R}^{q}$ are system states with $\mathcal{S} = \{\bm{s}| \underline{\bm{s}} \le \bm{s} \le \overline{\bm{s}}\}$ and $\mathcal{Y} = \{\bm{y}| \underline{\bm{y}} \le \bm{y} \le \overline{\bm{y}}\}$. Nonlinear vector fields  $f_1: \mathcal{S} \times \mathcal{Y} \rightarrow \mathbb{R}^{p}$, $f_2: \mathcal{S} \times \mathcal{Y} \rightarrow \mathbb{R}^{q}$ are Lipschitz continuous. Let $(\bm{s_e},\bm{y_e})$ be an equilibrium point of system \eqref{lemma_projection_dynamicsys}. Then, for any $\bm{s} \in \mathcal{S}$ and $\bm{y} \in \mathcal{Y}$, the following two inequalities hold
  \begin{subequations} \label{lemma_projection_inequal}%
    \begin{align}
      (\bm{s} - \bm{s_e})^{\top}[f_1(\bm{s},\bm{y})]_{\bm{s} - \underline{\bm{s}}}^{\overline{\bm{s}} - \bm{s}} &\le (\bm{s} - \bm{s_e})^{\top}f_1(\bm{s},\bm{y}) \label{lemma_projection_inequal1} \\
      (\bm{s} - \bm{s_e})^{\top}f_1(\bm{s_e},\bm{y_e}) &\le\bm{0}. \label{lemma_projection_inequal2}
    \end{align} 
  \end{subequations}
\end{lemma}
\textit{Proof:} By Lemma \ref{property_projection}, system dynamics \eqref{lemma_projection_dynamicsys1} and \eqref{lemma_projection_dynamicsys2} become 
\begin{subequations} \label{lemma_projection_eq1}%
  \begin{align}
    \dot{\bm{s}} &= [f_1(\bm{s},\bm{y})]_{\bm{s} - \underline{\bm{s}}}^{\overline{\bm{s}} - \bm{s}} = f_1(\bm{s},\bm{y}) - \bm{e_1} \label{lemma_projection_inequalnew1} \\
    \dot{\bm{y}} &= [f_2(\bm{s},\bm{y})]_{\bm{y} - \underline{\bm{y}}}^{\overline{\bm{y}} - \bm{y}} = f_2(\bm{s},\bm{y}) - \bm{e_2}, \label{lemma_projection_inequalnew2}
  \end{align} 
\end{subequations}
where $\bm{e_1} \in \mathcal{N}_{\mathcal{S}}(\bm{s}) = \{\bm{e_1}| \bm{e_1}^{\top}(\bm{s}' - \bm{s}) \le 0, \forall \bm{s}' \in \mathcal{S}\}$ and $\bm{e_2} \in \mathcal{N}_{\mathcal{Y}}(\bm{y}) = \{\bm{e_2}| \bm{e_2}^{\top}(\bm{y}' - \bm{y}) \le 0, \forall \bm{y}' \in \mathcal{Y}\}$. $\mathcal{N}_{\mathcal{S}}(\bm{s})$ (resp. $\mathcal{N}_{\mathcal{Y}}(\bm{y})$) denotes the normal cone to set $\mathcal{S}$ (resp. $\mathcal{Y}$) at $\bm{s}$ (resp. $\bm{y}$).

From \eqref{lemma_projection_inequalnew1}, we obtain
\begin{equation} \label{lemma_projection_eq2}
  \bm{e_1} = f_1(\bm{s},\bm{y}) - [f_1(\bm{s},\bm{y})]_{\bm{s} - \underline{\bm{s}}}^{\overline{\bm{s}} - \bm{s}}.
\end{equation}
By setting $\bm{s}' = \bm{s}_e$ in the definition of $\mathcal{N}_{\mathcal{S}}(\bm{s})$, we have 
\begin{equation}\label{eqa}
\bm{e}_1^{\top}(\bm{s}_e - \bm{s}) \le 0
\end{equation}  
Substituting \eqref{lemma_projection_eq2}  into \eqref{eqa} yields \eqref{lemma_projection_inequal1}.

Moreover, at the equilibrium point, \eqref{lemma_projection_inequalnew1} becomes
\begin{equation} \label{lemma_projection_inequalnew1_v2}%
  \dot{\bm{s}}_{\bm{e}} = [f_1(\bm{s_e},\bm{y_e})]_{\bm{s_e} - \underline{\bm{s}}}^{\overline{\bm{s}} - \bm{s_e}} = f_1(\bm{s_e},\bm{y_e}) - \bm{e_{1e}} = \bm{0}
\end{equation}
with $\bm{e_{1e}} \in \mathcal{N}_{\mathcal{S}}(\bm{s_e}) = \{\bm{e_{1e}}| \bm{e_{1e}}^{\top}(\bm{s}' - \bm{s_e}) \le 0, \forall \bm{s}' \in \mathcal{S}\}$. From \eqref{lemma_projection_inequalnew1_v2}, we have $\bm{e_{1e}} = f_1(\bm{s_e},\bm{y_e})$. Similar to \eqref{eqa}, we have $\bm{e_{1e}}^{\top}(\bm{s} - \bm{s_e}) \le 0$ by using the definition of $\mathcal{N}_{\mathcal{S}}(\bm{s_e})$. Substituting $\bm{e_{1e}} = f_1(\bm{s_e},\bm{y_e})$ into $\bm{e_{1e}}^{\top}(\bm{s} - \bm{s_e}) \le 0$ proves \eqref{lemma_projection_inequal2}. $\hfill\blacksquare$

\subsection{Set Invariance and Safety}
Beyond enforcing actuator limits, we must also ensure that critical state variables, e.g., system
frequency, remain within safe bounds during transients. This requires that the system state remains within a designated safety set for all time.

Consider system \eqref{PDS}, where the vector field is specified as follows
\begin{equation} \label{system_set_invariance}
  \dot{\bm{x}} = \left[F_0(\bm{x}) + F_1(\bm{x})\bm{u}(\bm{x})\right]^{\overline{\bm{x}} - \bm{x}}_{\bm{x} - \underline{\bm{x}}},
\end{equation}
where $F_{0}: \mathcal{X} \rightarrow \mathbb{R}^n$ and $F_{1}: \mathcal{X} \rightarrow \mathbb{R}^{n \times m}$ are Lipschitz continuous mappings, and $\bm{u}: \mathcal{X} \rightarrow \mathcal{U}$ is a Lipschitz continuous feedback controller. Here, $\mathcal{U} \subset \mathbb{R}^m$ is the set of all admissible control inputs.

Let the safety set $\mathcal{C} \subseteq \mathcal{X}$ be defined by 
\begin{equation}
  \mathcal{C} = \{\bm{x} \in \mathbb{R}^n | \phi_i(\bm{x}) \le 0, \forall i = 1,...,k\},
\end{equation}
where $\phi_i: \mathbb{R}^n \rightarrow \mathbb{R}$, $\forall i = 1,...,k$ are continuously differentiable functions.

\begin{definition}[Forward Invariance \cite{7937882}] \label{def_invar}
  A set $\mathcal{C}$ is forward invariant under \eqref{system_set_invariance} if every solution starting in $\mathcal{C}$ remains in $\mathcal{C}$ for all $t \in [0, T)$.
\end{definition}

In our context, ensuring the forward invariance of $\mathcal{C}$ corresponds to guaranteeing that the system frequency remains within its safety bounds at all times.

\subsection{Safety Control via Control Barrier Functions}
To enforce forward invariance of the safety set $\mathcal{C}$ through feedback control, we employ CBFs, which impose sufficient conditions to keep trajectories of system \eqref{system_set_invariance} within $\mathcal{C}$. The CBF conditions are
\begin{equation}\label{CBF-condition-g}
 \frac{\partial \phi_i(\bm{x})}{\partial \bm{x}}[F_0(\bm{x}) + F_1(\bm{x})\bm{u}(\bm{x})]^{\overline{\bm{x}} - \bm{x}}_{\bm{x} - \underline{\bm{x}}}  
    \le -\gamma_i \phi_i(\bm{x}), ~ \forall i = 1,...,k,
\end{equation}
where $\gamma_i > 0$ are controller parameters to be designed.

Based on this, we define the CBF as below.
\begin{definition}[CBF \cite{7937882}] \label{def_VCBF}
  Consider system \eqref{system_set_invariance}. The function $\phi(\bm{x}) = [\phi_1(\bm{x}),...,$ $\phi_k(\bm{x})]^{\top}$ is a CBF for the safety set $\mathcal{C}$ if there exists $\gamma_i > 0$, $\forall i = 1,...,k$, such that the set of admissible control inputs
  \begin{equation}
  \begin{aligned}
    \mathcal{K}(\bm{x}) = \{\bm{u} \in \mathcal{U} | \eqref{CBF-condition-g}~\text{holds}\}
  \end{aligned} \label{VCBF}%
\end{equation}
is nonempty for all $\bm{x} \in \mathcal{X}$.
\end{definition}

By using these results, the following lemma formalizes the safety guarantee.
\begin{lemma}[Forward Invariance of Set $\mathcal{C}$ via CBF \cite{7937882}] \label{lemma_safety}
  The set $\mathcal{C}$ is forward invariant if there exists a Lipschitz feedback controller $\bm{u} \in \mathcal{K}(\bm{x})$ for any $\bm{x} \in \mathcal{X}$.
\end{lemma}

In summary, projected dynamics provide a principled way to enforce input saturation (e.g., generation
capacity limits), while CBFs offer a formal method to enforce transient safety (e.g., frequency bounds). These tools will be combined in Section \ref{method} to construct a distributed FO-based controller that enforces both the capacity constraint and transient frequency safety.

\section{Distributed Frequency Control Method}\label{method}
Building on the tools introduced in Section \ref{preliminary}, we now present a distributed frequency control strategy that achieves the control objectives stated in Section \ref{formulation}. To achieve these goals in a modular manner, we propose a two-layer control framework for each control area. The first layer employs an FO-based controller to regulate frequency toward an optimal operation point, while the second layer utilizes a safety corrector to ensure the transient frequency safety. Particularly, the FO-based controller solves a global steady-state optimization problem by using a projected gradient-based algorithm to generate a reference control signal $P_{g,i}^r$. Then, the safety corrector uses the idea of CBFs to adjust the reference $P_{g,i}^r$ to the final control input $P_{g,i}$ by enforcing transient frequency safety and generation capacity constraints. 

The FO-based controller is implemented in a decentralized manner, while the safety corrector operates in a distributed fashion through information exchange between neighboring correctors. In the following, we describe the design of each module in detail, and then formulate the resulting closed-loop system.

\subsection{FO-Based Controller Design}
The first layer of the proposed control framework aims to dynamically drive the system toward a steady-state operating point that minimizes generation cost while restoring nominal frequency and scheduled tie-line power flows.

We begin by formulating the following convex  steady-state optimization problem that underlies the control objective and is commonly used in FO-based frequency control methods (e.g., \cite{Trip2016, Li2016, 7944568, Yang2021})
\begin{subequations}
  \begin{align}
    \min_{P_{g,i}}& \sum_{i \in \mathcal{N}}\frac{1}{2}a_{g,i} P_{g,i}^2 + b_{g,i} P_{g,i} \label{OSFC1_Problem_obj} \\
    \text{s.t.}~ & P_{g,i} - d_i = 0, \quad i \in \mathcal{N} \label{OSFC1_Problem_cons1} \\
    & \underline{P}_{g,i} \le P_{g,i} \le \overline{P}_{g,i}, \quad i \in \mathcal{N}. \label{OSFC1_Problem_cons3}
  \end{align} \label{OSFC1_Problem}%
\end{subequations}
The objective function \eqref{OSFC1_Problem_obj} is to minimize the total generation cost across the multi-area system. Constraint \eqref{OSFC1_Problem_cons1} ensures power balance. Constraint \eqref{OSFC1_Problem_cons3} is the generation capacity limit. In \eqref{OSFC1_Problem_obj}, $a_{g,i} \in \mathbb{R}_+$ and $b_{g,i} \in \mathbb{R}_+$ are quadratic and linear cost coefficients of the equivalent generating unit $i \in \mathcal{N}$, respectively, which are usually obtained via second-order polynomial fitting techniques (see \cite{Wood2014} for details). 

To solve problem \eqref{OSFC1_Problem}, we propose the following FO-based controller inspired by the projected gradient-based algorithm from \cite{9075378}
\begin{subequations}
  \begin{align}
    \dot{P}_{g,i}^{r} &= [-a_{g,i} P_{g,i}^{r} - b_{g,i} - \xi_i - w_i]_{P_{g,i}^r - \underline{P}_{g,i}}^{\overline{P}_{g,i} - P_{g,i}^r}, && i \in \mathcal{N} \label{OSFC1_FO1} \\
    \dot{\xi}_i &=  P_{g,i}^r - d_i, && i \in \mathcal{N}, \label{OSFC1_FO2} 
  \end{align} \label{OSFC1_FO}%
\end{subequations}
where $P_{g,i}^{r}$ is the generation reference signals for control area $i \in \mathcal{N}$; $\xi_i$ is the  Lagrange multiplier associated with constraint \eqref{OSFC1_Problem_cons1}. 

\begin{remark}
  Clearly, based on \eqref{OSFC1_FO}, each control area $i \in \mathcal{N}$ updates $P_{g,i}^r$ and $\xi_i$ independently, enabling a decentralized implementation of the algorithm. If $P_{g,i}^{r}$ is directly applied to the system by setting $P_{g,i} = P_{g,i}^r$ in \eqref{eq:PowerSysOG_Dynamicsb}, then the proposed controller \eqref{OSFC1_FO} can regulate system frequency and net tie-line powers while driving the system to the optimal solution of problem \eqref{OSFC1_Problem}. Thus, it satisfies requirement \textit{(i)}. Moreover, the projection operator in \eqref{OSFC1_FO1} enforces the capacity limits during both transients and steady-state operation, thereby addressing requirement \textit{(iii)}. The latter stands in contrast to many existing FO-based approaches, which often overlook capacity constraints during transients or only satisfy them at steady state (e.g., \cite{Trip2016,Trip2018,9869334,Li2016,7944568,Ste2017,Yang2021,10319778,zhao2024distributed,wang2020distributed}).  
\end{remark}

However, as with prior FO-based methods, transient frequency deviations may still fall outside the predefined safety bounds under large or fast disturbances. This motivates the introduction of a safety correction layer, addressed next.

\subsection{Safety Corrector Design}
As mentioned above, the FO-based controller steers the system toward an optimal steady state and respects capacity limits, but it does not guarantee that frequency trajectories remain within safe regions during transients. In low-inertia systems with high renewable penetration, frequency violations become more frequent due to the lower frequency nadirs and larger RoCoF. Such violations may also trigger protection schemes and lead to cascading failures \cite{8450880}. To address this, we introduce a safety corrector that dynamically modifies the generation reference $P_{g,i}^r$ to ensure the transient frequency safety. 

Define the frequency safety set $\mathcal{W}_i$ for each control area $i \in \mathcal{N}$ as 
\begin{equation} 
  \mathcal{W}_i = \left\{w_i ~| ~\underline{w}_i \le w_i \le \overline{w}_i \right\}.
\end{equation}
By defining $\underline{h}_i(w_i) = \underline{w}_i - w_i$ and $\overline{h}_i(w_i) = w_i - \overline{w}_i$,  $\mathcal{W}_i$ can be rewritten as 
\begin{equation}\label{safe_set}
  \mathcal{W}_i = \left\{ w_i ~|~\underline{h}_i(w_i) \le 0, \overline{h}_i(w_i) \le 0 \right\}.
\end{equation}

The design goal is to ensure
\begin{itemize}
  \item \textit{(ii.a)} Forward invariance of $\mathcal{W}_i$, $\forall i \in \mathcal{N}$, i.e., if $w_i(0) \in \mathcal{W}_i$, $\forall i \in \mathcal{N}$, then $w_i(t) \in \mathcal{W}_i$, $\forall i \in \mathcal{N}$, for all $t \ge 0$.
  \item \textit{(ii.b)} If $w_i(0) \notin \mathcal{W}_i$, then $w_i(t)$ enters $\mathcal{W}_i$ in finite time.
\end{itemize}

Based on \eqref{CBF-condition-g} and \eqref{safe_set}, the  CBF conditions are
\begin{align}
 \frac{\partial \underline{h}_i(w_i)}{\partial w_i}\dot{w}_i &\le -\beta_i \underline{h}_i(w_i), && \forall i \in \mathcal{N}\label{CBF-condition-a}\\
        \frac{\partial \overline{h}_i(w_i)}{\partial w_i}\dot{w}_i &\le -\beta_i \overline{h}_i(w_i), &&\forall i \in \mathcal{N},\label{CBF-condition-b}
\end{align}
where $\beta_i \in \mathbb{R}_+$, $\forall i \in \mathcal{N}$, are design parameters used to shape the enforcement strength of the CBF conditions. A smaller $\beta_i$ leads to a more conservative safety margin in $\mathcal{W}_i$, and vice versa. By the definitions of $\underline{h}_i(w_i)$ and $\overline{h}_i(w_i)$, \eqref{eq:PowerSysOG_Dynamicsb}, CBF conditions \eqref{CBF-condition-a} and \eqref{CBF-condition-b} are equivalent to
\begin{align}
  P_{g,i} &\ge \tilde{l}_{b,i}(w_i, \bar{\bm\alpha}_i)\label{OSFC1_Safety_cons3}\\
  P_{g,i} &\le \tilde{u}_{b,i}(w_i, \bar{\bm\alpha}_i), \label{OSFC1_Safety_cons4}
\end{align}
where the vector $\bar{\bm\alpha}_i$ consists of the reduced angle $\alpha_i$ of control area $i$ and all its neighbors' reduced angles $\alpha_j$, $\forall j\in\mathcal{N}_i$; $\tilde{l}_{b,i}(w_i, \bar{\bm\alpha}_i)$ and $ \tilde{u}_{b,i}(w_i, \bar{\bm\alpha}_i)$ are defined as 
\begin{equation}\label{l&u}
  \begin{aligned}
    \tilde{l}_{b,i}(w_i, \bar{\bm\alpha}_i) &= \underline{g}_i(w_i) + \varphi_i(\bar{\bm\alpha}_i) \\
    \tilde{u}_{b,i}(w_i, \bar{\bm\alpha}_i) &= \overline{g}_i(w_i) + \varphi_i(\bar{\bm\alpha}_i)
  \end{aligned}
\end{equation}
with $\underline{g}_i(w_i) = D_i w_i + d_i + \beta_i M_i (\underline{w}_{i} - w_i)$, $\overline{g}_i(w_i) = D_i w_i + d_i - \beta_i M_i (w_i - \overline{w}_{i})$, and $\varphi_i(\bar{\bm\alpha}_i) = \sum_{j \in \mathcal{N}_i} B_{ij}(\alpha_i - \alpha_j)$. 

Then, the corresponding admissible control input sets  $\mathcal{K}_i$, $\forall i \in \mathcal{N}$ are defined as
\begin{equation} \label{CBF_condition}
    \mathcal{K}_i= \left\{ P_{g,i} \in \mathcal{P}_i \left| \eqref{OSFC1_Safety_cons3},~ \eqref{OSFC1_Safety_cons4}
    \right.~\text{hold}
      \right\},
\end{equation}
with $\mathcal{P}_i = \{P_{g,i}| \underline{P}_{g,i} \le P_{g,i} \le \overline{P}_{g,i}\}$, $\forall i \in \mathcal{N}$ being the capacity limit sets. 

Based on Lemma~\ref{lemma_safety} in Section \ref{preliminary}.3, if the control inputs ${P}_{g,i}$, $\forall i \in \mathcal{N}$, lie within the admissible sets defined in~\eqref{CBF_condition}, i.e., ${P}_{g,i} \in \mathcal{K}_i$, then the safety sets $\mathcal{W}_i$ are forward invariant. 

Hence, if the reference inputs $P_{g,i}^r$ satisfy $P_{g,i}^r \in \mathcal{K}_i$, then we can directly set ${P}_{g,i} = P_{g,i}^r$. Otherwise, $P_{g,i}^r$ must be modified to ensure ${P}_{g,i} \in \mathcal{K}_i$. On the other hand, this modification should be as small as possible to preserve the optimality of the overall system.

Accordingly, the safety correction can be formulated as the following quadratic program:
\begin{subequations}
  \begin{align}
    \underset{P_{g,i}}{\min}&  ~\frac{1}{2}(P_{g,i} - P_{g,i}^r)^2 \label{OSFC1_Safety_obj} \\
    \text{s.t.} &~\underline{P}_{g,i} \le P_{g,i} \le \overline{P}_{g,i} \label{OSFC1_Safety_cons2} \\
    & \eqref{OSFC1_Safety_cons3}, \eqref{OSFC1_Safety_cons4}. \label{OSFC1_Safety_cons5}
  \end{align} \label{OSFC1_Safety}%
\end{subequations}
The objective \eqref{OSFC1_Safety_obj} is to minimize the deviation between $P_{g,i}^r$ and $P_{g,i}$. Constraints \eqref{OSFC1_Safety_cons2} and \eqref{OSFC1_Safety_cons5} enforce the CBF conditions derived from \eqref{CBF_condition}, guaranteeing  forward invariance of $\mathcal{W}_i$.  In the next section, we will prove that it ensures both \textit{(ii.a)} and \textit{(ii.b)}, and hence satisfies the overall requirement \textit{(ii)}. 

To ensure transient frequency safety, the final control inputs $P_{g,i}$ may differ from $P_{g,i}^r \in \mathcal{P}_i$, and thus may fall outside the capacity limit sets $\mathcal{P}_i$. The constraint \eqref{OSFC1_Safety_cons2} is imposed to ensure that $P_{g,i}$ satisfies the capacity constraint specified in requirement~\textit{(iii)}.

Problem \eqref{OSFC1_Safety} is a special quadratic program, which has analytical solutions \cite{boyd2004convex}, i.e.,
\begin{equation} \label{OSFC1_Safetynew}
  \begin{aligned}
    {P}_{g,i} =\min\left\{\max\left\{P_{g,i}^r, l_{b,i}(w_i,\bar{\bm\alpha}_i)\right\}, u_{b,i}(w_i,\bar{\bm\alpha}_i) \right\}
  \end{aligned}
\end{equation}
with $l_{b,i}(w_i,\bar{\bm\alpha}_i) = \max\{\underline{P}_{g,i}, \tilde{l}_{b,i}(w_i,\bar{\bm\alpha}_i)\}$ and $u_{b,i}(w_i,\bar{\bm\alpha}_i) = \min\{\overline{P}_{g,i}, \tilde{u}_{b,i}(w_i,\bar{\bm\alpha}_i)\}$. 

\begin{figure}[h]
  \centering
  \includegraphics[width=0.48\textwidth]{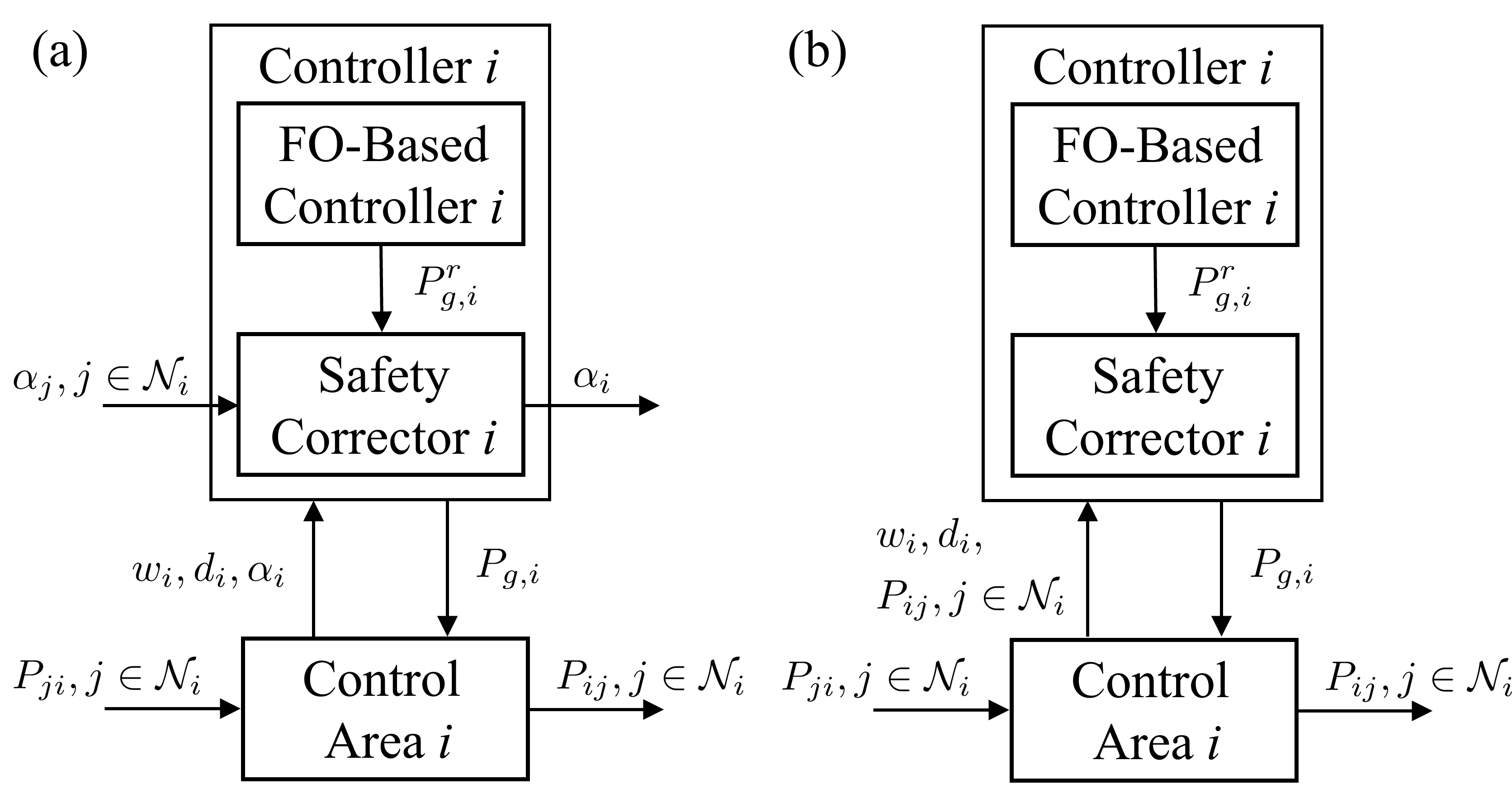}
  \caption{The architecture of the proposed control strategy. (a) Distributed implementation. (b) Decentralized implementation.}
  \label{Fig:OSFC}
\end{figure}

\begin{remark} \label{implementation_control}
The safety corrector \eqref{OSFC1_Safetynew} is distributed since each area $i \in \mathcal{N}$ only requires the net load $d_i$, local frequency $w_i$, its reduced phase angle $\alpha_i$ as well as its neighbors' reduced phase angles $\alpha_j$, $\forall j\in\mathcal{N}_i$.
 In practice, however, measuring the reduced angle $\alpha_i$ is generally impractical, even for a real SG, not to mention the equivalent generation unit. Fortunately, the terms $\underline{g}_i(w_i)$ and $\overline{g}_i(w_i)$ in $\tilde{l}_{b,i}(w_i, \bar{\bm\alpha}_i)$ and $\tilde{u}_{b,i}(w_i, \bar{\bm\alpha}_i)$ in \eqref{l&u} only depend on local information $\omega_i$ and $d_i$,  while $\varphi_i(\bar{\bm\alpha}_i)$ can be rewritten as  $\varphi_i(\bar{\bm\alpha}_i)=\sum_{j \in \mathcal{N}_i} P_{ij}$, where $P_{ij} = B_{ij}(\alpha_i - \alpha_j)$ is the power flow on line $(i,j)$. This relationship allows the dependence on reduced angle variables $\bar{\bm\alpha}_i$ to be replaced by tie-line flows $P_{ij}$ that can be measured locally. Thus, instead of using the unmeasurable $\alpha_i$, each area $i$ only needs to measure its tie-line power flows $P_{ij}$, $\forall j \in \mathcal{N}_i$. This also introduces another advantage of the proposed corrector, i.e., it can be implemented in a decentralized manner. Specifically, it can be computed only from local measurements, including net load $d_i$, frequency $w_i$, and tie-line flows $P_{ij}$. Hence, no global coordination or system-wide communication is required. Fig. \ref{Fig:OSFC}.~(a) and  (b) show the distributed and decentralized implementations of the proposed control method, respectively.
\end{remark}

\begin{remark} 
In contrast to existing distributed safety control approaches (e.g., \cite{ZHANG2021109335, 9031310}) that require real-time optimization, our method provides control signals in closed form, significantly reducing computational overhead. Moreover, it explicitly enforces transient capacity constraints—an aspect often overlooked by existing distributed control schemes with closed-form solutions (e.g., \cite{ZHANG2019274, YUAN2024105753}). Furthermore, unlike centralized safety-guaranteed frequency controllers (e.g., \cite{10005839}), the proposed distributed safety corrector is specifically designed for multi-area systems.  These features enhance the practicality and scalability of the proposed method for real-world deployment.
\end{remark}

\subsection{Closed-Loop System}
Define $\bm{\alpha}=[\alpha_1,...,\alpha_{N-1}]^\top \in \mathbb{R}^{N-1}$, $\bm{w_g}=[w_1,...,w_N]^\top \in \mathbb{R}^{N}$, $\bm{P}_{\bm{g}}^{\bm{r}} = [P_{g,1}^r,...,P_{g,N}^r]^{\top}$ $ \in \mathbb{R}^N$, $\bm{\xi} =[\xi_1,...,\xi_N]^\top\in \mathbb{R}^N$, $\bm{P_g} = [P_{g,1},...,P_{g,N}]^{\top}$ $\in \mathbb{R}^{N}$, and $\bm{d} = [d_1,...,d_N]^{\top} \in \mathbb{R}^{N}$. Then, the closed-loop system of \eqref{eq:PowerSysOG_Dynamics} with the designed FO-based controller \eqref{OSFC1_FO} and safety corrector \eqref{OSFC1_Safetynew} can be rewritten as 
\begin{subequations}
  \begin{align}
    \dot{\bm{\alpha}} =& \bm{T}\bm{w_g}  \label{closedloop3}\\
    \dot{\bm{w}}_{\bm{g}} =& \bm{M}^{-1}(-\bm{D}\bm{w_g} + \bm{P_g}(\bm{z}) - \bm{d} - \bm{F}\bm{B_l}\tilde{\bm{F}}^{\top}\bm{\alpha}) \label{closedloop4} \\
         \dot{\bm{P}}_{\bm{g}}^{\bm{r}} =& \left[-\bm{A_g}\bm{P}_{\bm{g}}^{\bm{r}} - \bm{b_g} - \bm{\xi} - \bm{w_g}\right]_{\bm{P_g^r} - \underline{\bm{P}}_{\bm{g}}}^{\overline{\bm{P}}_{\bm{g}} - \bm{P_g^r}} \label{closedloop1} \\
    \dot{\bm{\xi}} =&  \bm{P}_{\bm{g}}^{\bm{r}} - \bm{d} \label{closedloop2} \\
       \bm{P_g}(\bm{z}) =& \min\left\{\max\left\{\bm{P_g^r}, \bm{l_b}(\bm{w_g},\bm{\alpha})\right\}, \bm{u_b}(\bm{w_g},\bm{\alpha}) \right\}, \label{closedloop_corrector}
  \end{align} \label{closedloop}%
\end{subequations}
with $\bm{l_b}(\bm{w_g},\bm{\alpha}) = \max\{\underline{\bm{P}}_{\bm{g}}, \tilde{\bm{l}}_{\bm{b}}(\bm{w_g},\bm{\alpha})\}$, $\bm{u_b}(\bm{w_g},\bm{\alpha}) = \min\{\overline{\bm{P}}_{\bm{g}}, \tilde{\bm{u}}_{\bm{b}}(\bm{w_g},\bm{\alpha})\}$, $\tilde{\bm{l}}_{\bm{b}}(\bm{w_g},\bm{\alpha}) = \bm{D}\bm{w_g} + \bm{d} + \bm{F}\bm{B_l}\tilde{\bm{F}}^{\top}\bm{\alpha} + \bm{\Gamma}\bm{M}(\underline{\bm{w}}_{\bm{g}} - \bm{w_g})$, and $\tilde{\bm{u}}_{\bm{b}}(\bm{w_g},\bm{\alpha}) = \bm{D}\bm{w_g} + \bm{d} + \bm{F}\bm{B_l}\tilde{\bm{F}}^{\top}\bm{\alpha} - \bm{\Gamma}\bm{M}(\bm{w_g} - \overline{\bm{w}}_{\bm{g}})$.
Here $\bm{T} = [\bm{I}_{N-1},-\bm{1}_{N-1}] \in \mathbb{R}^{(N-1) \times N}$, $\bm{M} = \text{diag}\{M_1,...,M_N\} \in \mathbb{R}^{N \times N}$, $\bm{D} = \text{diag}\{D_1,...,D_N\} \in \mathbb{R}^{N \times N}$, $\bm{B_l} = \text{diag}\{...,B_{ij},...\} \in \mathbb{R}^{M \times M}$, $\bm{A_g} = \text{diag}\{\bm{a_g}\} \in \mathbb{R}^{N \times N}$ with $\bm{a_g} = [a_{g,1},...,a_{g,N}]^{\top}$ $ \in \mathbb{R}^{N}$, and  $\bm{\Gamma} = \text{diag}\{\beta_1,...,\beta_N\} \in \mathbb{R}^{N \times N}$ are constant matrices; $\bm{b_g} = [b_{g,1},...,b_{g,N}]^{\top} \in \mathbb{R}^{N}$, $\underline{\bm{P}}_{\bm{g}} = [\underline{P}_{g,1},...,\underline{P}_{g,N}]^{\top} \in \mathbb{R}^{N}$, $\overline{\bm{P}}_{\bm{g}} = [\overline{P}_{g,1},...,\overline{P}_{g,N}]^{\top} \in \mathbb{R}^{N}$,  $\underline{\bm{w}}_{\bm{g}} = [\underline{w}_1,...,\underline{w}_N]^{\top} \in \mathbb{R}^{N}$, and $\overline{\bm{w}}_{\bm{g}} = [\overline{w}_1,...,\overline{w}_N]^{\top} \in \mathbb{R}^{N}$ are constant vectors. $\bm{F} \in \mathbb{R}^{N \times M}$ is the incidence matrix of the connected undirected graph $\mathcal{G} = (\mathcal{N}, \mathcal{M})$ that modeling the topology of connected multi-area systems (cf. \cite{FB-LNS}). $\tilde{\bm{F}} \in \mathbb{R}^{(N-1) \times M}$ is the reduced incidence matrix obtained by removing the last row of $\bm{F}$. 
\begin{remark}
  Unlike RL- and MPC-based controllers (e.g., \cite{Yan2020,Zhao2022,10696981,Liu2023,Yang2021b}), which are computationally intensive or require offline training, our method features closed-form updates and enables real-time implementation with low computational overhead. Most existing FO-based methods (e.g., \cite{Trip2016,Trip2018,9869334,Li2016,7944568,Ste2017,Yang2021,10319778,zhao2024distributed,wang2020distributed}) either overlook transient frequency safety or enforce capacity constraints only at steady state. To solve this problem, we integrate the FO-based controller and safety corrector to simultaneously enforce the steady-state economic optimality (requirement~\textit{(i)}), the transient frequency safety (requirement~\textit{(ii)}), and the capacity constraint (requirement~\textit{(iii)}) within a unified control framework. This significantly enhances the practicality of the proposed method and distinguishes it from prior FO-based approaches. In addition, both control layers operate in a fully decentralized manner, relying solely on local measurements and neighbor-to-neighbor communication, thus satisfying requirement~\textit{(iv)}.
\end{remark}

\section{Closed-Loop Performance Analysis}\label{analysis}
This section presents a comprehensive analysis of the closed-loop system \eqref{closedloop}. We begin by establishing the existence and uniqueness of its solutions. We then demonstrate the optimality of the closed-loop system by showing that its equilibrium point contains the optimal solution of the optimization problem \eqref{OSFC1_Problem}. Finally, we prove the asymptotic stability of the closed-loop system and verify that the proposed control scheme ensures transient frequency safety.

\subsection{Existence and Uniqueness}
Define $\bm{z} = [\bm{\alpha}^{\top}, \bm{w_{g}}^{\top}, \bm{P}_{\bm{g}}^{\bm{r} \top}, \bm{\xi}^{\top}]^{\top}$, and let $\underline{\bm{z}}$, $\overline{\bm{z}}$ denote the vectors of lower and upper limits of $\bm{z}$, respectively. Define the set $\mathcal{Z}$ as $\mathcal{Z} = \{\bm{z}| \underline{\bm{z}} \le \bm{z} \le \overline{\bm{z}} \} \subseteq \mathbb{R}^{N-1} \times \mathbb{R}^N \times \mathcal{P} \times \mathbb{R}^N$ with $\mathcal{P} = \mathcal{P}_1 \times\cdots \times \mathcal{P}_N$. Clearly, the set $\mathcal{Z}$  is closed and convex.  Then, the closed-loop system \eqref{closedloop} is rewritten in the form of projected dynamical systems
\begin{equation} \label{closedloopnew}
\dot{\bm{z}} = [F(\bm{z})]^{\overline{\bm{z}} - \bm{z}}_{\bm{z} - \underline{\bm{z}}},
\end{equation}
where vector field $F(\bm{z})$ is defined as 
\begin{equation}
  F(\bm{z}) = \left[ 
  \begin{aligned}
    & \bm{T}\bm{w_g} \\
    & \bm{M}^{-1}(-\bm{D}\bm{w_g} + \bm{P_g}(\bm{z}) - \bm{d} - \bm{F}\bm{B_l}\tilde{\bm{F}}^{\top}\bm{\alpha}) \\
    & -\bm{A_g}\bm{P}_{\bm{g}}^{\bm{r}} - \bm{b_g} - \bm{\xi} - \bm{w_g} \\
    & \bm{P}_{\bm{g}}^{\bm{r}} - \bm{d} 
  \end{aligned} 
  \right],
\end{equation}
and $\bm{P_g}(\bm{z})$ is defined in \eqref{closedloop_corrector}.

Next, we will use \eqref{closedloopnew} to prove the existence and uniqueness of the closed-loop system. We first present the following assumption and lemma.

\begin{assumption} \label{assump_corrector_feasible}
  The optimization problem \eqref{OSFC1_Safety}  has at least one feasible solution for any $t\ge 0$.
\end{assumption}

\begin{remark}
  The feasibility of \eqref{OSFC1_Safety} depends on each control area having sufficient generation capacity to maintain transient frequency within safe bounds. This requirement aligns with standard operational practices in modern power systems, where adequate generation reserves are maintained to counteract frequency deviations resulting from sudden changes in net loads, e.g., in \cite{kundur1994,8450880,Wood2014}. Therefore, Assumption~\ref{assump_corrector_feasible} is mild and practically reasonable. Notably, this assumption is equivalent to that the safety corrector \eqref{OSFC1_Safetynew}, or equivalently, \eqref{OSFC1_Safety} has solutions for any $t\ge 0$.
\end{remark}

\begin{lemma}
\label{lemma_lipmaxmin}
  Let $g: \mathbb{R}^n \rightarrow \mathbb{R}$ and $h: \mathbb{R}^m \rightarrow \mathbb{R}$ be Lipschitz continuous functions with Lipschitz constants $L_g$ and $L_h$, respectively. Then, the following inequalities hold for any $\bm{x}, \bm{x'} \in \mathbb{R}^n$ and $\bm{y}, \bm{y'} \in \mathbb{R}^m$ with $L = \max\{L_g,L_h\}$
  \begin{subequations}\label{lemma_lipmaxmin_ineq}%
    \begin{align} 
      \|\max\{g(\bm{x}), h(\bm{y})\} &- \max\{g(\bm{x'}), h(\bm{y'})\}\| \nonumber \\
      &\le L( \|\bm{x}-\bm{x'}\| + \|\bm{y}-\bm{y'}\| ) \label{lemma_lipmax} \\ 
      \|\min\{g(\bm{x}), h(\bm{y})\} &- \min\{g(\bm{x'}), h(\bm{y'})\}\| \nonumber \\
      &\le L( \|\bm{x}-\bm{x'}\| + \|\bm{y}-\bm{y'}\| ). \label{lemma_lipmin} 
    \end{align}
  \end{subequations}
\end{lemma}
\textit{Proof:} To prove inequality \eqref{lemma_lipmax}, we first consider the case $g(\bm{x}) > h(\bm{y}) > g(\bm{x'}) > h(\bm{y'})$. Then, we have $\|\max\{g(\bm{x}), h(\bm{y})\} - \max\{g(\bm{x'}), h(\bm{y'})\}\| = \|g(\bm{x}) - g(\bm{x'})\|\le \max \{\|g(\bm{x}) - g(\bm{x'})\|, \|h(\bm{y}) - h(\bm{y'})\|\}$. Similarly, we can prove the following inequality holds 
\begin{equation} \label{lemma_lipmax2}
  \begin{aligned}
    \|\max\{g(\bm{x}), &h(\bm{y})\} - \max\{g(\bm{x'}), h(\bm{y'})\}\| \\
    &\le \max \{\|g(\bm{x}) - g(\bm{x'})\|, \|h(\bm{y}) - h(\bm{y'})\|\}
  \end{aligned}
\end{equation}
for all other possible orderings of $g(\bm{x})$, $h(\bm{y})$, $g(\bm{x'})$, and $h(\bm{y'})$.
 Combining \eqref{lemma_lipmax2} with the  Lipschitz property of $g(\bm{x})$ and $h(\bm{y})$ gives \eqref{lemma_lipmax}. The proof of inequality in \eqref{lemma_lipmin} follows a similar way, and thus is omitted.  $\hfill\blacksquare$

Based on these, we establish the following result.
\begin{theorem}
\label{theorem_existence}
Suppose Assumption \ref{assump_corrector_feasible} holds. Then, system \eqref{closedloopnew} has a unique Carathéodory solution $\bm{z}: [0, T) \rightarrow \mathcal{Z}$ for any initial conditions $\bm{z}(0) = \bm{z_0} \in \mathcal{Z}$ and some $T > 0$.
\end{theorem}
\textit{Proof:}  Based on Lemma \ref{PDS_existence}, to prove the existence and uniqueness of the solution of \eqref{closedloopnew}, we only need to establish that $F(\bm{z})$ is Lipschitz on $\mathcal{Z}$. Except for the safety corrector $\bm{P_g}(\bm{z})$, all other components of $F(\bm{z})$ are either affine or constant functions, and thus Lipschitz. Now, we will prove that $\bm{P_g}(\bm{z})$ is also Lipschitz on $\mathcal{Z}$.

The functions $\tilde{l}_{b,i}(w_i, \bar{\bm\alpha}_i)$ and $ \tilde{u}_{b,i}(w_i, \bar{\bm\alpha}_i)$ in \eqref{l&u} can be rewritten as 
\begin{equation}
  \begin{aligned}
    \tilde{l}_{b,i}(w_i, \bar{\bm\alpha}_i) =& D_i w_i + d_i + [\bm{F}\bm{B_l}\tilde{\bm{F}}^{\top}]_i\bm{\alpha} + \beta_i M_i(\underline{w}_{i} - w_i) \\
    =&\tilde{l}_{b,i}(\bm{w_g},\bm{\alpha}) \\
     \tilde{u}_{b,i}(w_i, \bar{\bm\alpha}_i) =& D_i w_i + d_i + [\bm{F}\bm{B_l}\tilde{\bm{F}}^{\top}]_i\bm{\alpha} - \beta_i M_i(w_i - \overline{w}_{i}) \\ 
     =&\tilde{u}_{b,i}(\bm{w_g},\bm{\alpha}),
  \end{aligned}
\end{equation}
which are the $i$-th entries of $\tilde{\bm{l}}_{\bm{b}}(\bm{w_g},\bm{\alpha})$ and $\tilde{\bm{u}}_{\bm{b}}(\bm{w_g},\bm{\alpha})$. It is clear that $\tilde{l}_{b,i}(\bm{w_g},\bm{\alpha})$ and $\tilde{u}_{b,i}(\bm{w_g},\bm{\alpha})$ are Lipschitz, i.e., 
\begin{subequations} \label{corrector_lipA_lubt}%
  \begin{align} 
    &\|\Delta \tilde{l}_{b,i}\| \le L_{A,i}(\|\Delta w_i\| + \|\Delta \bm{\alpha}\|) \label{corrector_lipA_lbt} \\
    &\|\Delta \tilde{u}_{b,i}\| \le L_{A,i}(\|\Delta w_i\| + \|\Delta \bm{\alpha}\|),  \label{corrector_lipA_ubt}
  \end{align}
\end{subequations}
for any $\bm{w_g}, \bm{w_g'} \in \mathbb{R}^{N}$ and $\bm{\alpha}, \bm{\alpha'} \in \mathbb{R}^{N-1}$, with $\Delta \tilde{l}_{b,i} = \tilde{l}_{b,i}(\bm{w_g},\bm{\alpha}) - \tilde{l}_{b,i}(\bm{w_g'},\bm{\alpha'})$, $\Delta \tilde{u}_{b,i} = \tilde{u}_{b,i}(\bm{w_g},\bm{\alpha}) - \tilde{u}_{b,i}(\bm{w_g'},\bm{\alpha'})$, $\Delta w_i = w_i - w_i'$, $\Delta \bm{\alpha} = \bm{\alpha} - \bm{\alpha}'$, and the Lipschitz constant $L_{A,i} = \max\{\|D_i - \beta_i M_i\|, \|[\bm{F}\bm{B_l}\tilde{\bm{F}}^{\top}]_i\|\}$.

Let $l_{b,i}(\bm{w_g},\bm{\alpha}) \in \mathbb{R}$ and $u_{b,i}(\bm{w_g},\bm{\alpha}) \in \mathbb{R}$, $\forall i\in\mathcal{N}$ be the $i$-th entries of $\bm{l}_{\bm{b}}(\bm{w_g},\bm{\alpha})$ and $\bm{u}_{\bm{b}}(\bm{w_g},\bm{\alpha})$, respectively, i.e.,
\begin{equation*}
  \begin{aligned}
    l_{b,i}(\bm{w_g},\bm{\alpha}) =& \max\{\underline{P}_{g,i}, \tilde{l}_{b,i}(\bm{w_g},\bm{\alpha})\} \\
    u_{b,i}(\bm{w_g},\bm{\alpha}) =& \min\{\overline{P}_{g,i}, \tilde{u}_{b,i}(\bm{w_g},\bm{\alpha})\}.
  \end{aligned}
\end{equation*}
Based on Lemma \ref{lemma_lipmaxmin}, $l_{b,i}(\bm{w_g},\bm{\alpha})$ and $u_{b,i}(\bm{w_g},\bm{\alpha})$ are also Lipschitz continuous,  i.e., 
\begin{subequations}
  \begin{align}
    &\|\Delta l_{b,i}\| \le L_{A,i}(\|\Delta w_i\| + \|\Delta \bm{\alpha}\|) \label{corrector_lipA_lb} \\
    &\|\Delta u_{b,i}\| \le L_{A,i}(\|\Delta w_i\| + \|\Delta \bm{\alpha}\|),  \label{corrector_lipA_ub}
  \end{align}
\end{subequations}
for any $\bm{w_g}, \bm{w_g'} \in \mathbb{R}^{N}$ and $\bm{\alpha}, \bm{\alpha'} \in \mathbb{R}^{N-1}$ with $\Delta l_{b,i} = l_{b,i}(\bm{w_g},\bm{\alpha}) - l_{b,i}(\bm{w_g'},\bm{\alpha'})$ and $\Delta u_{b,i} = u_{b,i}(\bm{w_g},\bm{\alpha}) - u_{b,i}(\bm{w_g'},\bm{\alpha'})$.

Define $\bm{p}(\bm{P_g^r},\bm{w_g},\bm{\alpha}) = \max\{\bm{P_g^r}, \bm{l_b}(\bm{w_g},\bm{\alpha})\}$ and let $p_i({P_{g,i}^r},\bm{w_g},\bm{\alpha}) = \max\{P_{g,i}^r, l_{b,i}(\bm{w_g},\bm{\alpha})\}$, $\forall i\in\mathcal{N}$ denote its $i$-th entry. Define $g(\bm{x})$ and $h(\bm{y})$ in Lemma \ref{lemma_lipmaxmin} as $g(\bm{x})=P_{g,i}^r$ and $h(\bm{y})=l_{b,i}(\bm{w_g},\bm{\alpha})$. Applying Lemma \ref{lemma_lipmaxmin} again gives $p_i({P_{g,i}^r},\bm{w_g},\bm{\alpha})$, $\forall i\in\mathcal{N}$ are Lipschitz, i.e., 
\begin{equation}
  \|\Delta p_i\| \le L_{B,i}(\|\Delta P_{g,i}^r\| + \|\Delta w_i \| + \|\Delta \bm{\alpha}\|),\label{corrector_lipA_p0} 
\end{equation}
for any ${P_{g,i}^r}, {P_{g,i}^{r'}} \in \mathcal{P}$ $\bm{w_g}, \bm{w_g'} \in \mathbb{R}^{N}$, and $\bm{\alpha}, \bm{\alpha'} \in \mathbb{R}^{N-1}$ with $\Delta p_i = p_i({P_{g,i}^r},\bm{w_g},\bm{\alpha}) - p_i({P_{g,i}^{r'}},\bm{w_g'},\bm{\alpha'})$, $\Delta P_{g,i}^r = P_{g,i}^r - P_{g,i}^{r'}$, and $L_{B,i} = \max \{1, L_{A,i}\}$.

The corrector \eqref{OSFC1_Safetynew} can be rewritten as ${P}_{g,i}(\bm{z}) =\min\left\{p_i({P_{g,i}^r},\bm{w_g},\bm{\alpha}), u_{b,i}(\bm{w_g},\bm{\alpha}) \right\}$. Since $p_i({P_{g,i}^r},\bm{w_g},\bm{\alpha})$ and $u_{b,i}(\bm{w_g},\bm{\alpha})$ are Lipschitz, by \eqref{lemma_lipmin} from Lemma \ref{lemma_lipmaxmin}, we conclude that for any $\bm{z}, \bm{z'} \in \mathcal{Z}$, there exists $L_{f,i} = \max\{1, \|D_i - \beta_i M_i\|, \|[\bm{F}\bm{B_l}\tilde{\bm{F}}^{\top}]_i\|\}$ such that
\begin{equation}
  \begin{aligned}
    \|P_{g,i}(\bm{z}) - P_{g,i}(\bm{z'})\| \le L_{f,i}(\|\Delta P_{g,i}^r\| + \|\Delta w_i\| &+ \|\Delta \bm{\alpha}\|), \\
    & \forall i\in\mathcal{N}.\label{corrector_lipA_p} 
  \end{aligned}
\end{equation}

Squaring both sides of \eqref{corrector_lipA_p} and summing over $i = 1,...,N$ yields
\begin{equation}
  \begin{aligned}
    &\|\bm{P_{g}}(\bm{z}) - \bm{P_{g}}(\bm{z'})\|^2 \\
    \le & \sum_{i = 1}^N L_{f,i}^2 (\|\Delta P_{g,i}^r\| + \|\Delta w_i\| + \|\Delta \bm{\alpha}\|)^2 \\
    \le & L_{f}^2 \sum_{i = 1}^N (\|\Delta P_{g,i}^r\| + \|\Delta w_i\| + \|\Delta \bm{\alpha}\|)^2 \\
    \le & 3L_{f}^2 \sum_{i = 1}^N (\|\Delta P_{g,i}^r\|^2 + \|\Delta w_i\|^2 + N\|\Delta \bm{\alpha}\|^2) \\
    \le & 3L_{f}^2 N \|\bm{z} - \bm{z'}\|^2,
  \end{aligned} \label{corrector_lip} 
\end{equation}
with $L_{f} = \max\{L_{f,1},...,L_{f,N}\}$, where the second inequality follows from $\sum_{i = 1}^N L_{f,i}^2 \le \sum_{i = 1}^N L_{f}^2$, and the third inequality is due to the fact $ab \le a^2 + b^2$ for any $a,b \in \mathbb{R}$. 

Taking the square root of both sides of \eqref{corrector_lip} gives $\|\bm{P_{g}}(\bm{z}) - \bm{P_{g}}(\bm{z'})\| \le \sqrt{3L_{f}^2 N} \|\bm{z} - \bm{z'}\|$ for any $\bm{z}, \bm{z'} \in \mathcal{Z}$, which proves that $\bm{P}_g(\bm{z})$ is Lipschitz continuous on $\mathcal{Z}$. Therefore, based on Lemma \ref{PDS_existence}, there exists a $T>0$,  the Carathéodory solution $\bm{z}(t)$ to system \eqref{closedloopnew} exists and is unique for all $t\in[0,T)$. $\hfill\blacksquare$

\begin{remark}
Theorem~\ref{theorem_existence} establishes the uniqueness and existence of solutions to system~\eqref{closedloopnew} over a time interval $[0, T)$ with some $T>0$. In Section \ref{analysis}.3, by using the invariance principle in Lemma \ref{PDS_invariance}, we will prove that this solution is complete,  i.e., $T \to \infty$.
\end{remark}
 
\subsection{Optimality}
For the optimality of the equilibrium point of the closed-loop system \eqref{closedloop}, we have the following result.
\begin{lemma}[Optimality] \label{lemma_optimality} Suppose Assumption \ref{assump_corrector_feasible} holds, and let $\bm{z_e} = [\bm{\alpha_e}^{\top}, \bm{w_{ge}}^{\top}, \bm{P}_{\bm{ge}}^{\bm{r} \top}, \bm{\xi_e}^{\top}]^{\top}$ be an equilibrium point of system \eqref{closedloop}.  Then,
  $(\bm{P}_{\bm{ge}}^{\bm{r}}, \bm{\xi_e})$ is the optimal solution of problem \eqref{OSFC1_Problem}, and vice versa. Moreover, the equilibrium point $\bm{z_e}$ is unique.
\end{lemma}
\textit{Proof:} To prove the optimality of  $(\bm{P}_{\bm{ge}}^{\bm{r}}, \bm{\xi_e})$, we first show that $\bm{w_{ge}} = \bm{0}$. Let $P_{g,i}(P_{g,e,i}^r, w_{e,i}, \bar{\bm\alpha}_{e,i})$ denote the $i$-th entry of $\bm{P}_g(\bm{z}_e)$ at the equilibrium point of \eqref{closedloop}. Substituting $\dot{\bm{w}}_{\bm{ge}} = \bm{0}$ into \eqref{closedloop4}  yields $\bm{P_{g}}(\bm{z_e}) = \bm{D}\bm{w_{ge}} + \bm{d} + \bm{F}\bm{B_l}\tilde{\bm{F}}^{\top}\bm{\alpha_e}$, or equivalently 
  \begin{equation} \label{pgz_eq}
    P_{g,i}(P_{ge,i}^r,w_{e,i},\bar{\bm\alpha}_{\bm{e}i}) = D_i w_{e,i} + d_i + [\bm{F}\bm{B_l}\tilde{\bm{F}}^{\top}]_i\bm{\alpha_e}, \forall i \in \mathcal{N}.
  \end{equation} 
By summing up \eqref{pgz_eq} for all $i \in \mathcal{N}$ and noting that $\sum_{i \in \mathcal{N}} [\bm{F}\bm{B_l}\tilde{\bm{F}}^{\top}]_i\bm{\alpha_e} = 0$, we have 
 \begin{equation}\label{lemma_eq_sum}
  \sum_{i \in \mathcal{N}} \left(P_{g,i}(P_{ge,i}^r,w_{e,i},\bar{\bm\alpha}_{\bm{e}i}) - d_i\right) - \sum_{i \in \mathcal{N}} D_i w_{e,i} = 0.
 \end{equation}
Further, substituting $\dot{\bm{\xi}}_{\bm{e}} = \bm{0}$ into \eqref{closedloop2} gives 
  \begin{equation}\label{Pr&d}
P_{ge,i}^r = d_i, \forall i \in \mathcal{N}. 
\end{equation}

Then, we claim that
\begin{equation} \label{Pg_eq}
  \begin{aligned}
    P_{g,i}(P_{ge,i}^r,w_{e,i},\bar{\bm\alpha}_{\bm{e}i}) =& \min\{\max\{P_{ge,i}^r, \underline{P}_{g,i}\}, \overline{P}_{g,i} \} \\
    =& P_{ge,i}^r,
  \end{aligned}
\end{equation}
where the last equality follows from $P_{ge,i}^r \in [\underline{P}_{g,i}, \overline{P}_{g,i}]$.  According to \eqref{OSFC1_Safetynew}, equality \eqref{Pg_eq} will hold if, for all $i \in \mathcal{N}$,
\begin{subequations} \label{lemma_eq_conclusion}%
  \begin{align}
    P_{g,i}(P_{ge,i}^r,w_{e,i},\bar{\bm\alpha}_{\bm{e}i}) &\ne \tilde{l}_{b,i}(w_{e,i}, \bar{\bm\alpha}_{\bm{e}i})  \label{lemma_eq_conclusion1}\\
    P_{g,i}(P_{ge,i}^r,w_{e,i},\bar{\bm\alpha}_{\bm{e}i}) &\ne \tilde{u}_{b,i}(w_{e,i}, \bar{\bm\alpha}_{\bm{e}i}),  \label{lemma_eq_conclusion2}
  \end{align}
\end{subequations}
except for a trivial case with $\tilde{l}_{b,i}(w_{e,i}, \bar{\bm\alpha}_{\bm{e}i})$ $= \underline{P}_{g,i}$ and $\tilde{u}_{b,i}(w_{e,i}, \bar{\bm\alpha}_{\bm{e}i}) = \overline{P}_{g,i}$. Under the trivial case, it is clear that \eqref{Pg_eq} holds.
Therefore, we will exclude the trivial case and prove \eqref{lemma_eq_conclusion} holds at the equilibrium point for all the other cases. 

Let us first prove \eqref{lemma_eq_conclusion1} by using contradiction. Suppose  \eqref{lemma_eq_conclusion1} does not hold and there exists a nonempty subset $\tilde{\mathcal{N}} \subseteq \mathcal{N}$ such that
\begin{equation}\label{lemma_eq_conclusion1neq}
P_{g,i}(P_{ge,i}^r,w_{e,i},\bar{\bm\alpha}_{\bm{e}i}) = \tilde{l}_{b,i}(w_{e,i} \bar{\bm\alpha}_{\bm{e}i}), \forall i\in\tilde{\mathcal{N}}.
\end{equation}
Substituting \eqref{l&u} into \eqref{lemma_eq_conclusion1neq} and comparing with \eqref{pgz_eq}, we obtain $\beta_i M_i (\underline{w}_{i} - w_{e,i}) = 0$, $ \forall i\in\tilde{\mathcal{N}}$. The positiveness of $\beta_i$ and $M_i$ gives $w_{e,i} = \underline{w}_i< 0$, $\forall i \in \tilde{\mathcal{N}}$. On the other hand, substituting $\dot{\bm{\alpha}}_{\bm{e}} = \bm{0}$ into \eqref{closedloop3} gives $w_{e,i} = w_{e,j}$, $\forall i, j \in \mathcal{N}$. Therefore, we have 
\begin{equation}\label{wNtilde}
w_{e,i} = w_{e,j} = \underline{w}_i< 0, \forall i, j \in \mathcal{N}. 
\end{equation}
If $\min \{\underline{w}_i, \forall i \in \mathcal{N} \}< \min \{\underline{w}_j, \forall j \in \mathcal{N} \setminus \tilde{\mathcal{N}}\}$, then we have $w_{e,i} \ne w_{e,j}$, $\forall i \in \tilde{\mathcal{N}}$, $\forall j \in \mathcal{N} \setminus \tilde{\mathcal{N}}$, which contradicts to \eqref{wNtilde} and gives \eqref{lemma_eq_conclusion1} holds. If instead $\min \{\underline{w}_i, \forall i \in \mathcal{N} \} \ge \min \{\underline{w}_j, \forall j \in \mathcal{N} \setminus \tilde{\mathcal{N}}\}$, then \eqref{wNtilde} still holds and we proceed with further analysis.

Combining \eqref{OSFC1_Safetynew} with \eqref{lemma_eq_conclusion1neq}, we have 
\begin{equation}\label{Pg&Pr}
P_{g,i}(P_{ge,i}^r,w_{e,i},\bar{\bm\alpha}_{\bm{e}i})  \ge P_{ge,i}^r, \forall i \in \tilde{\mathcal{N}},
\end{equation}
which together with \eqref{Pr&d} yields
  \begin{equation}\label{lemma_eq_sum_p1}
    \sum_{i \in \tilde{\mathcal{N}}} \left(P_{g,i}(P_{ge,i}^r,w_{e,i},\bar{\bm\alpha}_{\bm{e}i}) - d_i\right) \ge \sum_{i \in \tilde{\mathcal{N}}} \left(P_{ge,i}^r - d_i\right) = 0.
  \end{equation}
On the other hand, for any $i \in \mathcal{N} \setminus \tilde{\mathcal{N}}$, if $P_{g,i}(P_{ge,i}^r,w_{e,i},\bar{\bm\alpha}_{\bm{e}i}) = \tilde{u}_{b,i}(w_{e,i}, \bar{\bm\alpha}_{\bm{e}i})$, then we have $w_{e,i} = \overline{w}_i> 0$ by using a similar way as those to obtain \eqref{wNtilde}, which contradicts to \eqref{wNtilde}. Therefore, we have $P_{g,i}(P_{ge,i}^r,w_{e,i},\bar{\bm\alpha}_{\bm{e}i}) \ne \tilde{u}_{b,i}(w_{e,i}, \bar{\bm\alpha}_{\bm{e}i})$ and $P_{g,i}(P_{ge,i}^r,w_{e,i},\bar{\bm\alpha}_{\bm{e}i}) \ne \tilde{l}_{b,i}(w_{e,i}, \bar{\bm\alpha}_{\bm{e}i})$, $\forall i \in \mathcal{N} \setminus \tilde{\mathcal{N}}$, and thus have $P_{g,i}(P_{ge,i}^r,w_{e,i},\bar{\bm\alpha}_{\bm{e}i})  = P_{ge,i}^r$, which combining with \eqref{Pr&d} gives
  \begin{equation}\label{lemma_eq_sum_p2}
    \begin{aligned}
      \sum_{i \in \mathcal{N} \setminus \tilde{\mathcal{N}}} \left(P_{g,i}(P_{ge,i}^r,w_{e,i},\bar{\bm\alpha}_{\bm{e}i}) - d_i\right) =& \sum_{i \in \mathcal{N} \setminus \tilde{\mathcal{N}}} \left(P_{ge,i}^r - d_i\right) \\
      =& 0.
    \end{aligned}
  \end{equation}
Combining \eqref{lemma_eq_sum_p1} and \eqref{lemma_eq_sum_p2} gives $\sum_{i \in \mathcal{N}} \left(P_{g,i}(P_{ge,i}^r,w_{e,i},\bar{\bm\alpha}_{\bm{e}i})\right.$ $\left.- d_i\right) \ge 0$. This with $D_i > 0$ and $w_{e,i} <0$ gives
 \begin{equation}\label{lemma_eq_sum2}
  \sum_{i \in \mathcal{N}} \left(P_{g,i}(P_{ge,i}^r,w_{e,i},\bar{\bm\alpha}_{\bm{e}i}) - d_i\right) - \sum_{i \in \mathcal{N}} D_i w_{e,i} > 0,
 \end{equation}
 which contradicts to \eqref{lemma_eq_sum}. Thus, we have \eqref{lemma_eq_conclusion1} holds.

 Similarly, we can prove \eqref{lemma_eq_conclusion2} also holds.

Combining \eqref{lemma_eq_sum}  and \eqref{Pg_eq} gives 
\begin{equation} \label{eq_condition1}
  0 = -\sum_{i \in \mathcal{N}}D_i w_{e,i} + \sum_{i \in \mathcal{N}} P_{ge,i}^r - \sum_{i \in \mathcal{N}} d_i.
\end{equation}
Combining \eqref{eq_condition1}, $P_{ge,i}^r = d_i$, $D_i > 0, \forall i \in \mathcal{N}$, and $w_{e,i} = w_{e,j}, \forall i, j \in \mathcal{N}$ gives $w_{e,i} = w_{e,j} = 0, \forall i, j \in \mathcal{N}$, establishing $\bm{w_{ge}} = \bm{0}$. 

Substituting $\bm{w_{ge}} = \bm{0}$ and $\dot{\bm{P}}_{\bm{ge}}^{\bm{r}} = \bm{0}$ into \eqref{closedloop1} gives 
\begin{equation} \label{eq_condition7}
  \bm{0} = [-\bm{A_g}\bm{P}_{\bm{ge}}^{\bm{r}} - \bm{b_g} - \bm{\xi_e}]_{\bm{P_{ge}^r} - \underline{\bm{P}}_{\bm{g}}}^{\overline{\bm{P}}_{\bm{g}} - \bm{P_{ge}^r}}.
\end{equation}
Moreover, substituting $\dot{\bm{\xi}}_{\bm{e}} = \bm{0}$ into \eqref{closedloop2} gives
\begin{equation} \label{eq_condition5}
  \bm{0} = \bm{P}_{\bm{ge}}^{\bm{r}} - \bm{d}.
\end{equation}
By\cite[Theorem 3.2]{Cojocaru2004}, equations \eqref{eq_condition7} and \eqref{eq_condition5} are equivalent to the Karush-Kuhn-Tucker conditions of problem \eqref{OSFC1_Problem}. Therefore, $(\bm{P}_{\bm{ge}}^{\bm{r}}, \bm{\xi_e})$ in the equilibrium point of system \eqref{closedloop} is equivalent to the optimal solution of problem \eqref{OSFC1_Problem}.

Finally, we prove the uniqueness of $\bm{z_e}$. Since problem \eqref{OSFC1_Problem} is strictly convex, its optimal solution $\bm{P_g}^*$ is unique. The constraint matrix associated with \eqref{OSFC1_Problem_cons1} is the identity matrix $\bm{I}_N$, which has full row rank. As a result, the corresponding optimal Lagrange multiplier $\bm{\xi}^*$ is also unique \cite{boyd2004convex}. Since $(\bm{P}_{\bm{ge}}^{\bm{r}}, \bm{\xi_e})$ corresponds to the optimal solution $(\bm{P_g}^*,\bm{\xi}^*)$ of problem \eqref{OSFC1_Problem}, we conclude that $(\bm{P}_{\bm{ge}}^{\bm{r}}, \bm{\xi_e})$ is unique. Substituting $\dot{\bm{w}}_{\bm{ge}} = \bm{0}$, $\bm{w_{ge}} = \bm{0}$, and \eqref{Pg_eq} into \eqref{closedloop4} gives  
\begin{equation} \label{eq_condition6}
  \bm{0} = \bm{P}_{\bm{ge}}^{\bm{r}} - \bm{d} - \bm{F}\bm{B_l}\tilde{\bm{F}}^{\top}\bm{\alpha_e}.
\end{equation}
Since the matrix $\bm{F}\bm{B_l}\tilde{\bm{F}}^{\top}$ is full column rank \cite{FB-LNS}, the uniqueness of $\bm{P_{ge}}^{\bm{r}}$ ensures the uniqueness of $\bm{\alpha}_e$ for a given net load $\bm{d}$. Combining all these facts concludes the uniqueness. $\hfill\blacksquare$

\begin{remark}
  At the equilibrium point of \eqref{closedloop}, $\bm{w_{ge}} = \bm{0}$ ensures that the frequency regulation target in \eqref{steady_requirement1} is achieved. Combining this, $P_{ge,i}^r = d_i, \forall i \in \mathcal{N}$, and $P_{ge,i}^r = D_i w_{e,i} + d_i + [\bm{F}\bm{B_l}\tilde{\bm{F}}^{\top}]_i\bm{\alpha_e}, \forall i \in \mathcal{N}$ further gives $[\bm{F}\bm{B_l}\tilde{\bm{F}}^{\top}]_i\bm{\alpha_e} = 0, \forall i \in \mathcal{N}$. By the definition of net tie-line powers, this yields $[\bm{F}\bm{B_l}\tilde{\bm{F}}^{\top}]_i\bm{\alpha_e} = \sum_{j \in \mathcal{N}_i }B_{ij}(\alpha_{e,i} - \alpha_{e,j}) = P_{t,i}^s = 0, \forall i \in \mathcal{N}$, fulfilling the net tie-line powers regulation requirement in \eqref{steady_requirement2}.
\end{remark}

\begin{remark}
  Lemma~\ref{lemma_optimality} establishes that $(\bm{P}_{\bm{ge}}^{\bm{r}}, \bm{\xi_e})$ in the equilibrium point of \eqref{closedloop} solves problem~\eqref{OSFC1_Problem} exactly. Our framework explicitly addresses transient frequency safety and generation capacity constraints, outperforming the prior FO-based methods that neglect these requirements (see, e.g., \cite{Trip2016,Trip2018,9869334,Li2016,7944568,Ste2017,Yang2021,10319778,zhao2024distributed,wang2020distributed,10488734}). To this end, we introduce an explicit safety corrector \eqref{closedloop_corrector} that projects the reference signal $P_{g,i}^r$ onto the admissible set $\mathcal{K}_i$ for each area $i \in \mathcal{N}$ at all times. Crucially, this corrector is carefully constructed to guarantee $P_{g,i}(P_{ge,i}^r,w_{e,i},\bar{\bm\alpha}_{\bm{e}i}) = P_{ge,i}^r$, $\forall i \in \mathcal{N}$, thus ensuring that constraint satisfaction of CBF conditions does not compromise the steady-state optimality. This design bridges the gap between transient constraint enforcement and long-term economic operation in a unified manner.
\end{remark}

\subsection{Stability}
Based on the abovementioned results, we are ready to establish the stability result.
\begin{theorem} \label{theorem_stability}
 Suppose Assumption \ref{assump_corrector_feasible} holds.  The equilibrium point of system \eqref{closedloop} is asymptotically stable.
\end{theorem}
\textit{Proof:} Let $\hat{\bm{\alpha}} = \bm{\alpha} - \bm{\alpha_e}$, $\hat{\bm{w}}_{\bm{g}} = \bm{w_g} - \bm{w_{ge}}$, $\hat{\bm{P}}_{\bm{g}}^{\bm{r}} = \bm{P_g^r} - \bm{P_{ge}^r}$, $\hat{\bm{\xi}} = \bm{\xi} - \bm{\xi_e}$, and $\hat{\bm{z}} = \bm{z} - \bm{z_e}$. Define set $\hat{\mathcal{Z}} = \{\hat{\bm{z}} | \underline{\bm{P}}_{\bm{g}} - \bm{P_{ge}^r} \le \hat{\bm{P}}_{\bm{g}}^{\bm{r}} \le  \overline{\bm{P}}_{\bm{g}} - \bm{P_{ge}^r}\}$. Consider the following Lyapunov function candidate
\begin{equation} \label{theorem_lyapunov_func}
  V(\hat{\bm{z}}) = \frac{1}{2}\hat{\bm{\alpha}}^{\top}\tilde{\bm{F}}\bm{B_l}\tilde{\bm{F}}^{\top}\hat{\bm{\alpha}} + \frac{1}{2}\hat{\bm{w}}_{\bm{g}}^{\top}\bm{M}\hat{\bm{w}}_{\bm{g}} + \frac{1}{2}||\hat{\bm{P}}_{\bm{g}}^{\bm{r}}||^{2} + \frac{1}{2}||\hat{\bm{\xi}}||^{2}.
\end{equation}
Since the reduced incidence matrix $\tilde{\bm{F}}$ is full-row-rank \cite[Chapter 8]{FB-LNS}, combining this and the positive definite matrix $\bm{B_l}$ ($B_{ij} > 0$, $\forall (i,j) \in \mathcal{M}$) gives that matrix $\tilde{\bm{F}}\bm{B_l}\tilde{\bm{F}}^{\top}$ is positive definite \cite[Chapter 7]{Horn2013}. Thus, $V(\hat{\bm{z}}) \ge 0$ on the set $\hat{\mathcal{Z}}$ and $V(\hat{\bm{z}}) = 0$ if and only if $\hat{\bm{z}} = \bm{0}$.

Taking the derivative of $V(\hat{\bm{z}})$ along system \eqref{closedloop}, we have
\begin{equation} \label{theorem_dot_V}
  \begin{aligned}
    \dot{V}(\hat{\bm{z}}) =& \hat{\bm{\alpha}}^{\top}\tilde{\bm{F}}\bm{B_l}\tilde{\bm{F}}^{\top}\bm{T}\bm{w_g} + \hat{\bm{w}}_{\bm{g}}^{\top}(-\bm{D}\bm{w_g} + \bm{P_g}(\bm{z}) - \bm{d}  \\
    & - \bm{F}\bm{B_l}\tilde{\bm{F}}^{\top}\bm{\alpha}) + \hat{\bm{P}}_{\bm{g}}^{\bm{r}\top}[-\bm{A_g}\bm{P}_{\bm{g}}^{\bm{r}} - \bm{b_g} - \bm{\xi} \\
    &- \bm{w_g}]_{\bm{P_g^r} - \underline{\bm{P}}_{\bm{g}}}^{\overline{\bm{P}}_{\bm{g}} - \bm{P_g^r}} + \hat{\bm{\xi}}^{\top}(\bm{P_g^r} - \bm{d}).
  \end{aligned}
\end{equation}
For the first term on the right-hand side of \eqref{theorem_dot_V}, since $\dot{\bm{\alpha}}_{\bm{e}} = \bm{T}\bm{w_{ge}} = \bm{0}$, we have
\begin{equation} \label{theorem_dot_V_term3}
  \hat{\bm{\alpha}}^{\top}\tilde{\bm{F}}\bm{B_l}\tilde{\bm{F}}^{\top}\bm{T}\bm{w_g} = \hat{\bm{\alpha}}^{\top}\tilde{\bm{F}}\bm{B_l}\tilde{\bm{F}}^{\top}\bm{T}\hat{\bm{w}}_{\bm{g}}.
\end{equation}
For the second term on the right-hand side of \eqref{theorem_dot_V}, since $\bm{M}\dot{\bm{w}}_{\bm{ge}} = -\bm{D}\bm{w_{ge}} + \bm{P_g}(\bm{z_e}) - \bm{d} - \bm{F}\bm{B_l}\tilde{\bm{F}}^{\top}\bm{\alpha_e} = \bm{0}$, we have
\begin{equation} \label{theorem_dot_V_term4}
  \begin{aligned}
    &\hat{\bm{w}}_{\bm{g}}^{\top}(-\bm{D}\bm{w_g} + \bm{P_g}(\bm{z}) - \bm{d} - \bm{F}\bm{B_l}\tilde{\bm{F}}^{\top}\bm{\alpha}) \\
    =& \hat{\bm{w}}_{\bm{g}}^{\top}(-\bm{D}\hat{\bm{w}}_{\bm{g}} + \hat{\bm{P_g}}(\bm{z}) - \bm{F}\bm{B_l}\tilde{\bm{F}}^{\top}\hat{\bm{\alpha}} ),
  \end{aligned}
\end{equation}
where $\hat{\bm{P_g}}(\bm{z}) = \bm{P_g}(\bm{z}) - \bm{P_g}(\bm{z_e})$.

\noindent{For the third term on the right-hand side of \eqref{theorem_dot_V}, let $\bm{s} = \bm{P_g^r}$ and $\bm{y} = [\bm{\alpha}^{\top}, \bm{w_{g}}^{\top}, \bm{\xi}^{\top}]^{\top}$ in Lemma \ref{lemma_projection} and then we have}
\begin{equation} \label{theorem_dot_V_term1}
  \begin{aligned}
    &\hat{\bm{P}}_{\bm{g}}^{\bm{r}\top}[-\bm{A_g}\bm{P}_{\bm{g}}^{\bm{r}} - \bm{b_g} - \bm{\xi} - \bm{w_g}]_{\bm{P_g^r} - \underline{\bm{P}}_{\bm{g}}}^{\overline{\bm{P}}_{\bm{g}} - \bm{P_g^r}}\\
    \le& \hat{\bm{P}}_{\bm{g}}^{\bm{r}\top}(-\bm{A_g}\bm{P}_{\bm{g}}^{\bm{r}} - \bm{b_g} - \bm{\xi} - \bm{w_g}) - \hat{\bm{P}}_{\bm{g}}^{\bm{r}\top}(-\bm{A_g}\bm{P}_{\bm{ge}}^{\bm{r}} - \bm{b_g} \\
    &- \bm{\xi_e} - \bm{w_{ge}}) \\
    =& \hat{\bm{P}}_{\bm{g}}^{\bm{r}\top}(-\bm{A_g}\hat{\bm{P}}_{\bm{g}}^{\bm{r}} - \hat{\bm{\xi}} - \hat{\bm{w}}_{\bm{g}}).
  \end{aligned}
\end{equation}
For the fourth term on the right-hand side of \eqref{theorem_dot_V}, since $\dot{\bm{\xi}}_{\bm{e}} = \bm{P}_{\bm{ge}}^{\bm{r}} - \bm{d} = \bm{0}$, we have
\begin{equation} \label{theorem_dot_V_term2}
  \hat{\bm{\xi}}^{\top}(\bm{P_g^r} - \bm{d}) = \hat{\bm{\xi}}^{\top}(\bm{P_g^r} - \bm{d}  - \dot{\bm{\xi}}_{\bm{e}}) = \hat{\bm{\xi}}^{\top}\hat{\bm{P}}_{\bm{g}}^{\bm{r}}.
\end{equation}
Substituting \eqref{theorem_dot_V_term3}-\eqref{theorem_dot_V_term2} into \eqref{theorem_dot_V} gives 
\begin{equation} \label{theorem_dot_V2}
  \dot{V} \le -\hat{\bm{P}}_{\bm{g}}^{\bm{r}\top}\bm{A_g}\hat{\bm{P}}_{\bm{g}}^{\bm{r}} - \hat{\bm{w}}_{\bm{g}}^{\top}\bm{D}\hat{\bm{w}}_{\bm{g}} - \hat{\bm{P}}_{\bm{g}}^{\bm{r}\top} \hat{\bm{w}}_{\bm{g}} + \hat{\bm{w}}_{\bm{g}}^{\top}\hat{\bm{P_g}}(\bm{z}).
\end{equation}
Substituting $\bm{w_{ge}} = \bm{0}$ and \eqref{Pg_eq} into the last two terms of \eqref{theorem_dot_V2} yields $- \hat{\bm{P}}_{\bm{g}}^{\bm{r}\top} \hat{\bm{w}}_{\bm{g}} + \hat{\bm{w}}_{\bm{g}}^{\top}\hat{\bm{P_g}}(\bm{z}) = \bm{w_g}^{\top}(\bm{P_g}(\bm{z}) - \bm{P_g^r})$. Now, consider the two possible cases of $w_i$, $\forall i \in \mathcal{N}$.
\begin{itemize}
  \item $w_{i} \le 0$. According to constraints \eqref{OSFC1_Safety_cons2} and \eqref{OSFC1_Safety_cons3}, we have $P_{g,i}(\bm{z}) \ge l_{b,i}(\bm{w_g},\bm{\alpha})$. If $P_{g,i}^r > l_{b,i}(\bm{w_g},\bm{\alpha})$, then $P_{g,i}(\bm{z}) = P_{g,i}^r$, and thus $w_{i}(P_{g,i}(\bm{z}) - P_{g,i}^r) = 0$. If $P_{g,i}^r \le l_{b,i}(\bm{w_g},\bm{\alpha})$, then $P_{g,i}(\bm{z}) = l_{b,i}(\bm{w_g},\bm{\alpha})$, and thus $w_{i}(P_{g,i}(\bm{z}) - P_{g,i}^r) = w_{i}(l_{b,i}(\bm{w_g},\bm{\alpha}) - P_{g,i}^r) \le 0$. Therefore, $w_{i}(P_{g,i}(\bm{z}) - P_{g,i}^r) \le 0$ holds. 
  \item $w_{i} > 0$. By following the similar analysis as the first case, we conclude that $w_{i}(P_{g,i}(\bm{z}) - P_{g,i}^r) \le 0$.
\end{itemize}
Combining these two facts gives $\bm{w_g}^{\top}(\bm{P_g}(\bm{z}) - \bm{P_g^r}) \le \bm{0}$. Thus, \eqref{theorem_dot_V2} becomes
\begin{equation} \label{dotV_less0}
  \begin{aligned}
    \dot{V}(\hat{\bm{z}}) &\le -\hat{\bm{P}}_{\bm{g}}^{\bm{r}\top}\bm{A_g}\hat{\bm{P}}_{\bm{g}}^{\bm{r}} - \hat{\bm{w}}_{\bm{g}}^{\top}\bm{D}\hat{\bm{w}}_{\bm{g}} \\
    &\le 0 
  \end{aligned}
\end{equation}
with positive definite matrices $\bm{A_g}$ and $\bm{D}$ ($a_{g,i} > 0$ and $D_i > 0, \forall i \in \mathcal{N}$).

Define $\hat{\mathcal{Z}}_1 = \{\hat{\bm{z}} \in \hat{\mathcal{Z}} | V(\hat{\bm{z}}) \le V(\hat{\bm{z}}(0))\}$. The non-negative quadratic terms in \eqref{theorem_lyapunov_func} ensures that $\hat{\bm{\alpha}}$, $\hat{\bm{w}}_{\bm{g}}$, $\hat{\bm{P}}_{\bm{g}}^{\bm{r}}$, $\hat{\bm{\xi}}$ are bounded in $\hat{\mathcal{Z}}_1$. Moreover, \eqref{dotV_less0} ensures that $\hat{\bm{z}}$ remains in $\hat{\mathcal{Z}}_1$ for any time $t \ge 0$. Therefore, the set $\hat{\mathcal{Z}}_1$ is a compact forward invariant set for system \eqref{closedloop}.

Let $\hat{\mathcal{Z}}_2$ be the largest weakly invariant set satisfying $\hat{\mathcal{Z}}_2 \subset \mathcal{E} = \text{cl}\{\hat{\bm{z}} \in \hat{\mathcal{Z}}_1 | \dot{V}(\hat{\bm{z}}) = 0\}$. Next, we prove that the set $\hat{\mathcal{Z}}_2$ only contains the point $\hat{\bm{z}} = \bm{0}$. Since matrices $\bm{A_g}$ and $\bm{D}$ are positive definite, combining $\dot{V}(\hat{\bm{z}}) = \bm{0}$ and \eqref{dotV_less0} gives $\hat{\bm{P}}_{\bm{g}}^{\bm{r}} = \bm{0}$ and $\hat{\bm{w}}_{\bm{g}} = \bm{0}$, and thus $\bm{P}_{\bm{g}}^{\bm{r}} = \bm{P}_{\bm{ge}}^{\bm{r}}$ and $\bm{w}_{\bm{g}} = \bm{w}_{\bm{ge}} = \bm{0}$. Substituting $\bm{w}_{\bm{ge}} = \bm{0}$ into \eqref{closedloop3} gives $\dot{\bm{\alpha}} = \bm{0}$ holds at the set $\hat{\mathcal{Z}}_2$. Finally, combining $\bm{P}_{\bm{g}}^{\bm{r}} = \bm{P}_{\bm{ge}}^{\bm{r}}$ and \eqref{closedloop2} gives $\dot{\hat{\bm{\xi}}} = \dot{\bm{\xi}} =  \bm{P}_{\bm{ge}}^{\bm{r}} - \bm{d} = \bm{0}$ holds at the set $\hat{\mathcal{Z}}_2$. Thereofer, according to Lemma \ref{PDS_invariance}, every solution to \eqref{closedloop} starting from $\hat{\bm{z}}(0) \in \hat{\mathcal{Z}}_1$ is complete and asymptotically converge to the largest weakly invariant set $\hat{\mathcal{Z}}_2$, i.e., $\lim_{t \rightarrow \infty} \hat{\bm{z}}(t)= \bm{0}$. Thus, we have $\lim_{t \rightarrow \infty} \bm{z}(t)= \bm{z_e}$. Considering the uniqueness of $\bm{z_e}$ from Lemma \ref{lemma_optimality}, the equilibrium point $\bm{z_e}$ of \eqref{closedloop} is asymptotically
stable. $\hfill\blacksquare$

\begin{remark}
In practice, the net load $d_i$ may be obtained either through direct measurements or short-term predictions \cite{8066371,1458194}. When timely measurements are available, the proposed controller uses the measured values directly. However, due to the rapid variability of $d_i$—especially under high penetration of renewables—real-time measurements may not always be accessible. In such cases, the controller operates based on predicted net loads, under the assumption that these predictions are accurate enough. This is a reasonable assumption. For the short-term net load predictions (e.g., over subseconds to a few seconds in our paper), net load variations are generally small, allowing for high-accuracy predictions using established forecasting techniques (e.g., \cite{1458194,10454597}). Nevertheless, as demonstrated in the simulations in Section \ref{case}, the proposed method remains robust even in the presence of moderate prediction errors in the net load. In summary, $d_i$ refers to either the measured or predicted value throughout the paper, depending on availability.
\end{remark}

\begin{remark}
In Theorem~\ref{theorem_stability}, we establish the asymptotic stability of the closed-loop system under step changes in net loads. However, in practical systems, net loads are often time-varying due to the inherent variability of renewable generation. A more general stability analysis in this setting may be conducted using the framework of input-to-state stability \cite{Khalil2002}. This represents a promising direction for future work. Nonetheless, the proposed method remains effective in dealing with time-varying net loads, as demonstrated by simulation results in Section \ref{case}.
\end{remark}

\begin{remark}
The stability result in Theorem~\ref{theorem_stability} is derived under the assumption that the equivalent parameters $M_i$ and $D_i$ are known. In practice, however, these parameters may vary over time due to changes in generation configurations and load characteristics. Although their exact values may not be available in real-time, simulation results in Section \ref{case} demonstrate that the proposed method remains effective under moderate uncertainties in $M_i$ and $D_i$. A rigorous theoretical analysis of robustness with respect to such parameter variations will be our future work.
\end{remark}

\subsection{Transient Frequency Safety}
Based on the aforementioned discussions, we are ready to establish the transient frequency safety.

\begin{theorem}[Transient Frequency Safety] \label{theorem_safety}
  Consider the closed-loop system \eqref{closedloop}. Suppose Assumption \ref{assump_corrector_feasible} holds. Then, for any control area $i \in \mathcal{N}$,
  
  \textit{(\ref{theorem_safety}.a)} If $w_i(0) \in [\underline{w}_i, \overline{w}_i]$ at $t = 0$, then $ w_i(t) \in [\underline{w}_i, \overline{w}_i]$ for any $t > 0$.
  
  \textit{(\ref{theorem_safety}.b)} If $w_i(0) \notin [\underline{w}_i, \overline{w}_i]$ at $t = 0$, then $ w_i(t)$ approaches $[\underline{w}_i, \overline{w}_i]$ monotonically. Besides, there exists a finite time $t_1 > 0$ such that $w_i(t) \in [\underline{w}_i, \overline{w}_i]$ for any $t \ge t_1$.
\end{theorem}
\textit{Proof:} Given Assumption \ref{assump_corrector_feasible}, combining the Lipschitz corrector $\bm{P_g}(\bm{z})$ by Theorem \ref{theorem_existence} and Lemma \ref{lemma_safety} gives that the sets $\mathcal{W}_i$, $\forall i \in \mathcal{N}$ are forward invariant, which ensures \textit{(\ref{theorem_safety}.a)} holds.

To prove the monotonicity in \textit{(\ref{theorem_safety}.b)}, we need to prove that for any control area $i \in \mathcal{N}$, $\dot{w}_i (t) \le 0$ when $w_i(t) \ge \overline{w}_i$ and $\dot{w}_i (t) \ge 0$ when $w_i(t) \le \underline{w}_i$. We begin with the first case. 

When $w_i(t) \ge \overline{w}_i$, since $\beta_i > 0$ and $M_i > 0$, we have $\beta_i M_i (w_i - \overline{w}_i) \ge 0$ holds. Combining this and $\tilde{u}_{b,i}(\bm{w_g}, \bm{\alpha}) = D_i w_i + d_i + [\bm{F}\bm{B_l}\tilde{\bm{F}}^{\top}]_i\bm{\alpha} - \beta_i M_i(w_i - \overline{w}_i)$ gives that 
\begin{equation} \label{ud_tilde_less0}
  \begin{aligned}
    &\tilde{u}_{b,i}(\bm{w_g}, \bm{\alpha}) - D_i w_i - d_i - [\bm{F}\bm{B_l}\tilde{\bm{F}}^{\top}]_i\bm{\alpha} \\
    =& - \beta_i M_i(w_i - \overline{w}_i) \\
    \le& 0.
  \end{aligned}
\end{equation}
Moreover, according to \eqref{OSFC1_Safety_cons4}, we have $P_{g,i}(\bm{z}) \le \tilde{u}_{b,i}(\bm{w_g}, \bm{\alpha})$. Combining this, \eqref{ud_tilde_less0}, and \eqref{closedloop4}, we have 
\begin{equation}
  \begin{aligned}
    M_i \dot{w}_i =& P_{g,i}(\bm{z}) - D_i w_i - d_i - [\bm{F}\bm{B_l}\tilde{\bm{F}}^{\top}]_i\bm{\alpha} \\
    \le& \tilde{u}_{b,i}(\bm{w_g}, \bm{\alpha}) - D_i w_i - d_i - [\bm{F}\bm{B_l}\tilde{\bm{F}}^{\top}]_i\bm{\alpha} \\
    \le& 0.
  \end{aligned}
\end{equation}
Since $M_i > 0$, we have $\dot{w}_i(t) \le 0$ holds. Then, by following the similar analysis, we conclude that $\dot{w}_i (t) \ge 0$ when $w_i(t) \le \underline{w}_i$. Combining these two facts ensures the monotonicity of $w_i(t)$ towards $[\underline{w}_i, \overline{w}_i]$. Finally, since $\lim_{t \rightarrow \infty} w_i(t) = w_{e,i} = 0$ with $w_{e,i} \in [\underline{w}_i, \overline{w}_i]$ as proved in Theorem \ref{theorem_stability}, there exists a finite time $t_1$ such that $w_i(t_1) \in [\underline{w}_i, \overline{w}_i]$. By utilizing the forward invariance of set $\mathcal{W}$, we have $w_i(t) \in [\underline{w}_i, \overline{w}_i]$ for any $t \ge t_1$. Thus, we conclude that \textit{(\ref{theorem_safety}.b)} holds. $\hfill\blacksquare$
\begin{remark}
  Theorem~\ref{theorem_safety} does not provide a theoretical upper bound on the finite time $t_1$ when $w_i(0) \notin [\underline{w}_i, \overline{w}_i]$. However, by establishing a lower bound on $\dot{w}_i$, one could derive an explicit upper bound on $t_1$. Such a result would be valuable for enhancing the practical applicability of the proposed control strategy, particularly in time-critical scenarios. This direction will be the focus of future research. Nonetheless, as shown in Section \ref{case}, the proposed frequency control strategy brings the frequency back within the safety limits in approximately 3 seconds, which is sufficiently fast for practical use.
\end{remark}

\section{Case Study}\label{case}
\begin{figure}[H]
  \centering
  \includegraphics[width=0.13\textwidth]{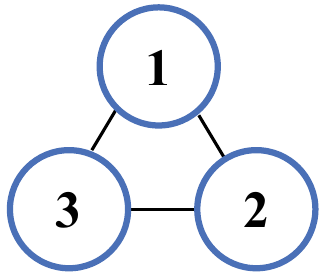}
  \caption{Connection structure of the 3-area test system.}
  \label{Fig:MultiArea}
\end{figure}

In this section, we test the effectiveness of the proposed control method on a 3-area power system with a power base 100 MW and the nominal frequency 50 Hz. The system topology is shown in Fig.~\ref{Fig:MultiArea}, and the system parameters are adopted from \cite{kundur1994}. In accordance with the operational guidelines of power systems (e.g., \cite{ENTSOE_FrequencyQuality_2015}), the safe frequency region is selected as [49.9, 50.1] Hz.

\begin{figure*} 
  \centering
  \includegraphics[width=0.8\textwidth]{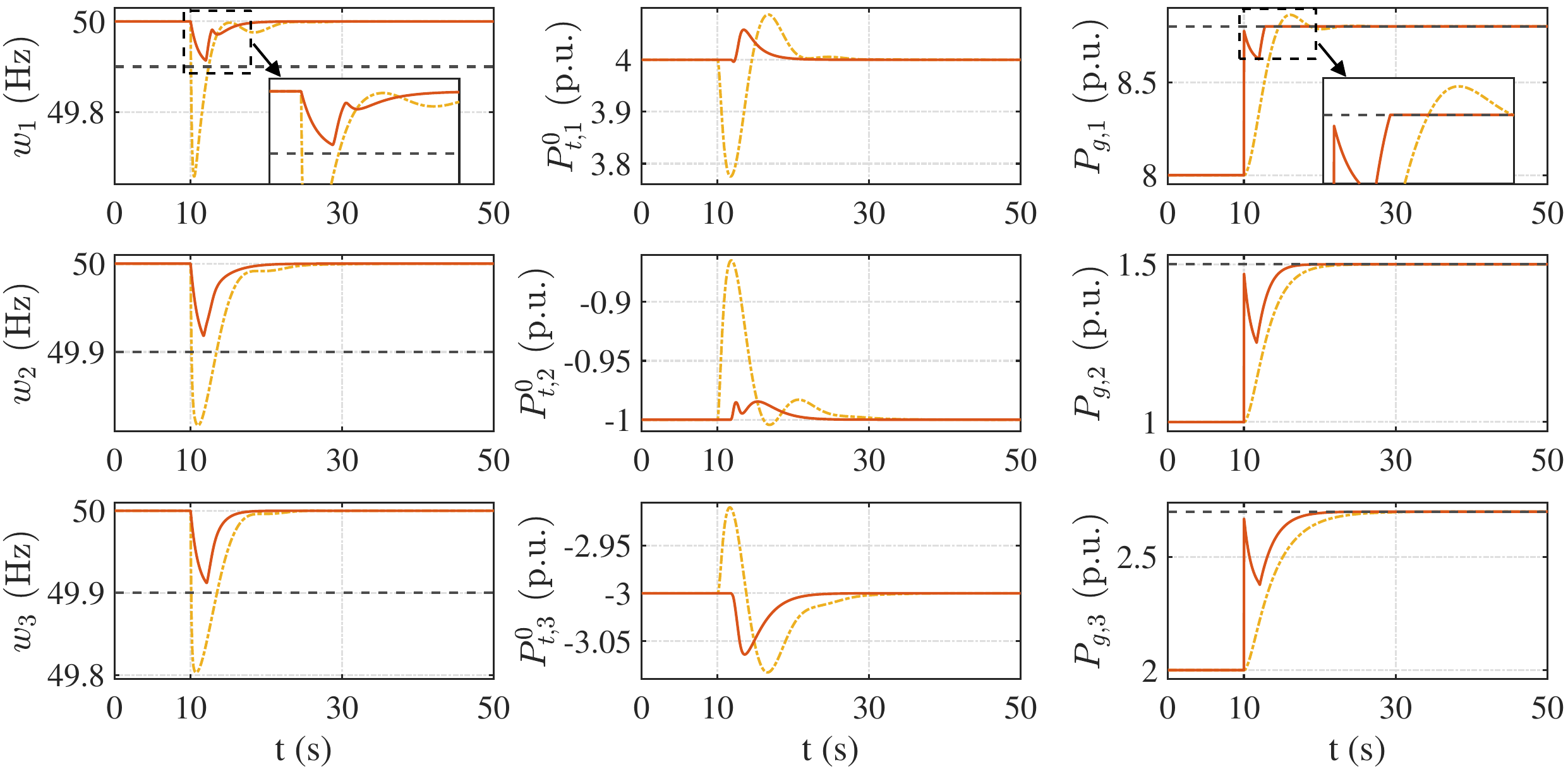}
  \caption{Control performance comparison under step net load changes, where yellow and red lines denote the responses under SFC and the proposed method, respectively. Left column: frequency responses of areas 1, 2, and 3 (black dashed lines denote lower frequency limits). Middle column: net tie-line powers of areas 1, 2, and 3. Right column: generation outputs of areas 1, 2, and 3 (black dashed lines denote upper generation capacity limits).}
  \label{Fig:Compare_StepChange}
\end{figure*}

To simulate the power system dynamics, we adopt the nonlinear swing equation \eqref{eq:PowerSys_Swing} (cf. \cite{Wood2014}) instead of the linearized system \eqref{eq:PowerSysOG_Dynamics} used in deriving the proposed method.
\begin{subequations}
  \begin{align}
     \dot{\bm{\alpha}} =& \bm{T}\bm{w_g}  \label{eq:PowerSys_Swinga}\\
     \bm{M}\dot{\bm{w}}_{\bm{g}} =& -\bm{D}\bm{w_g} + \bm{P_g} - \bm{d} - \bm{F}\hat{\bm{B}}_{\bm{l}}\sin (\tilde{\bm{F}}^{\top}\bm{\alpha}), \label{eq:PowerSys_Swingb} 
  \end{align}
  \label{eq:PowerSys_Swing}%
\end{subequations}
where $\hat{\bm{B}}_{\bm{l}} = \text{diag}\{...,\hat{B}_{ij},...\} \in \mathbb{R}^{M \times M}$ with susceptance $\hat{B}_{ij}$ of line $(i,j) \in \mathcal{M}$ as the diagonal entry.

Simulations are conducted by MATLAB on a laptop equipped with an Intel Core i7-11800H CPU and 16 GB RAM. The problem \eqref{OSFC1_Problem} is solved using MOSEK \cite{mosek} and is used as a benchmark to assess steady-state optimality. To further demonstrate the advantages of the proposed method, we compare its control performance against that of the secondary frequency control (SFC) approach in \cite{Ste2017}. SFC is a distributed FO-based frequency control method, which restores frequency and net tie-line powers and guarantees steady-state optimality, but does not address the transient frequency safety or the transient capacity constraint.

Moreover, we also compare these two methods with the conventional AGC. However, due to its significantly inferior performance in all evaluated aspects, the AGC results are omitted from this paper.

\subsection{Step Net Load Changes}

Firstly, we evaluate the performance of these two methods under the step net load changes. At $t = 10$ s, step changes of 0.8, 0.5, and 0.7 p.u. are applied to the net loads of areas 1, 2, and 3, respectively. 

Fig.~\ref{Fig:Compare_StepChange} illustrates the resulting system frequency responses, net tie-line powers, and generation outputs. Both methods successfully restore frequency and net tie-line powers to their nominal values. However, the proposed method achieves faster recovery of frequency and net tie-line powers, maintains frequency within its safety region ($w_i \in [49.9, 50.1]$ Hz, $i = 1, 2, 3$) and satisfies the transient capacity constraint ($P_{g,1} \in [7.2, 8.8]$ p.u., $P_{g,2} \in [0.5, 1.5]$ p.u., and $P_{g,3} \in [1.3, 2.7]$ p.u.), outperforming SFC which does not enforce these constraints. Furthermore, the cost trajectory of the proposed method converges to the optimal value (\$ 8599), consistent with the benchmark, thereby confirming its steady-state optimality.

\begin{figure*}
  \centering
  \includegraphics[width=0.75\textwidth]{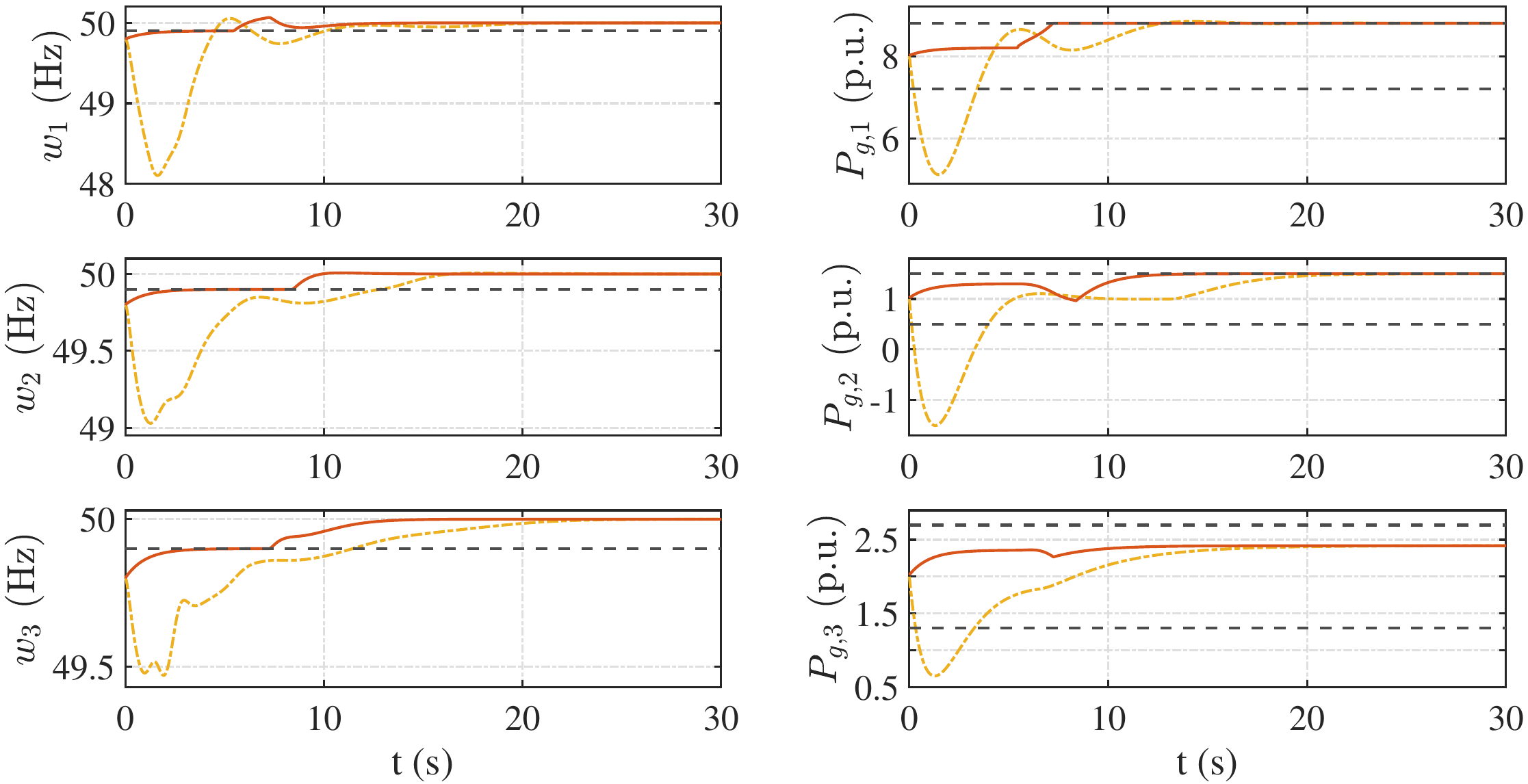}
  \caption{Frequency restoration comparison of various control methods, where yellow and red lines denote the responses under SFC and the proposed method, respectively. Left column: frequency responses of areas 1, 2, and 3 (black dashed lines denote lower frequency limits). Right column: generation outputs of areas 1, 2, and 3 (black dashed lines denote admissible output ranges).}
  \label{Fig:Compare_frequency}
\end{figure*}

We further evaluate the performance of different methods when the initial frequency is outside the safe region, specifically $w_i = 49.8$ Hz, $\forall i \in \mathcal{N}$. At $t = 0$ s, SFC and the proposed method are activated to restore system frequency. The simulation results, presented in Fig. \ref{Fig:Compare_frequency}, show that the proposed method not only achieves a faster frequency recovery (around 3 s) compared to SFC but also ensures the capacity constraints at all times. Furthermore, once the frequency enters the safe region [49.9, 50.1] Hz, the proposed method ensures it remains within bounds throughout the transient procedure.

\subsection{Time-Varying Net Load Changes}
\begin{figure}[H]
  \centering
  \includegraphics[width=0.45\textwidth]{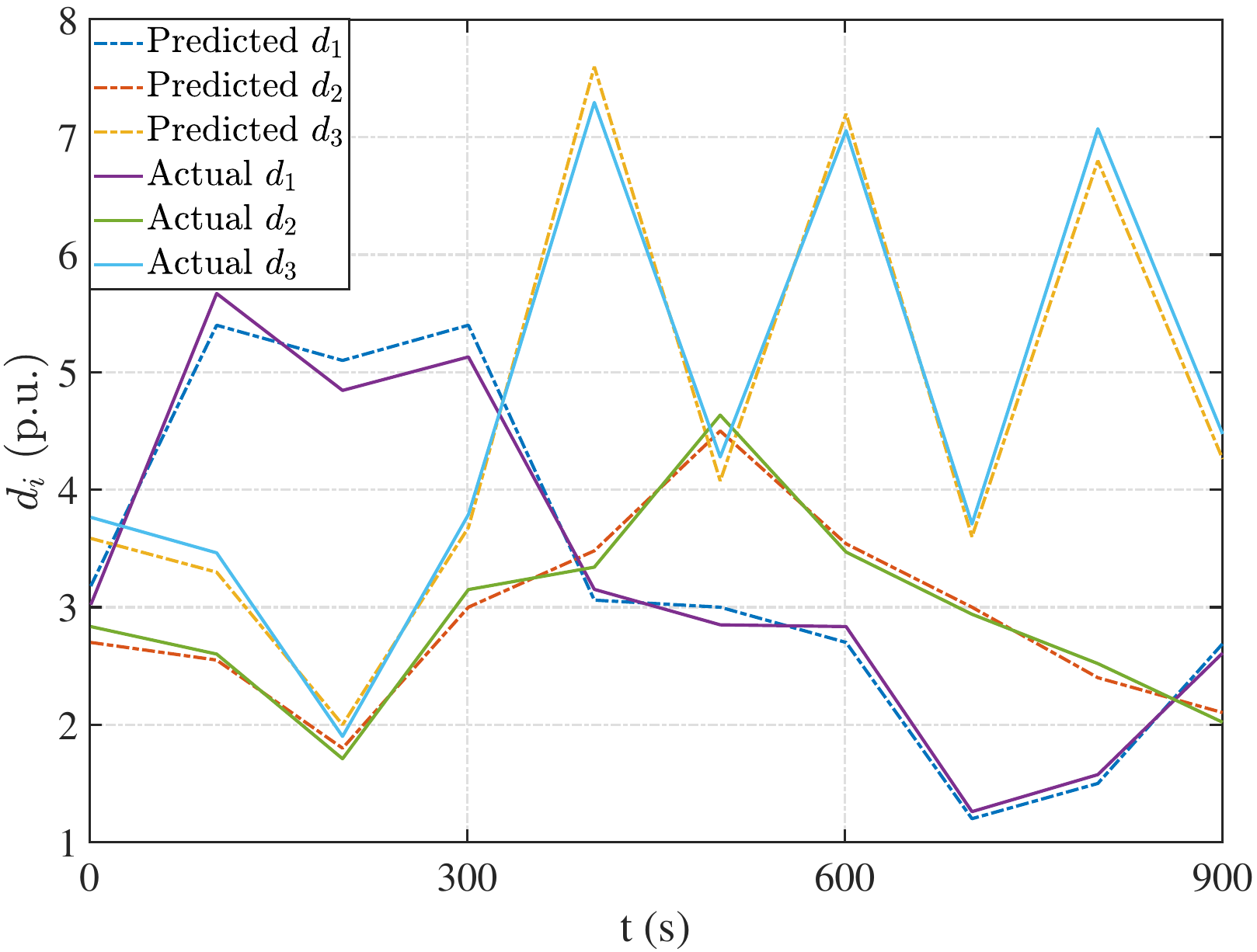}
  \caption{Time-varying total net load.}
  \label{Fig:Net_Load}
\end{figure}
We further evaluate the performance of the proposed control method under time-varying net load conditions over a 900-second simulation period, as illustrated in Fig.~\ref{Fig:Net_Load}. Specifically, we introduce a 5\% prediction error in the net loads $d_i$, $i = 1, 2, 3$, which represents a sufficiently large deviation as reported in the literature (e.g., \cite{1458194,10454597}). To simulate variable damping, we allow each $D_i$ to fluctuate within $\pm 5\%$, which also reflects sufficiently large operational variations in power systems \cite{6320655}. Thirdly, we simulate changes in rotational inertia $M_i$, as described below.
\begin{itemize}
  \item At $t = 300$ s, $M_1$ is decreased to 80 \% of its original value.
  \item At $t = 450$ s, $M_2$ is increased to 150 \% of its original value.
  \item At $t = 600$ s, $M_3$ is decreased to 60 \% of its original value.
\end{itemize}

The simulation setup reflects extreme operating conditions that rarely occur in practical power systems. For example, despite the volatility of renewables, area net load rarely increases by nearly 100\% within 100 seconds \cite{1458194}, as in area 3 from 300 s to 400 s in Fig. \ref{Fig:Net_Load}. The goal of this setting is to test the effectiveness and robustness of the proposed method under such challenging conditions.

\begin{figure*}
  \centering
  \includegraphics[width=0.8\textwidth]{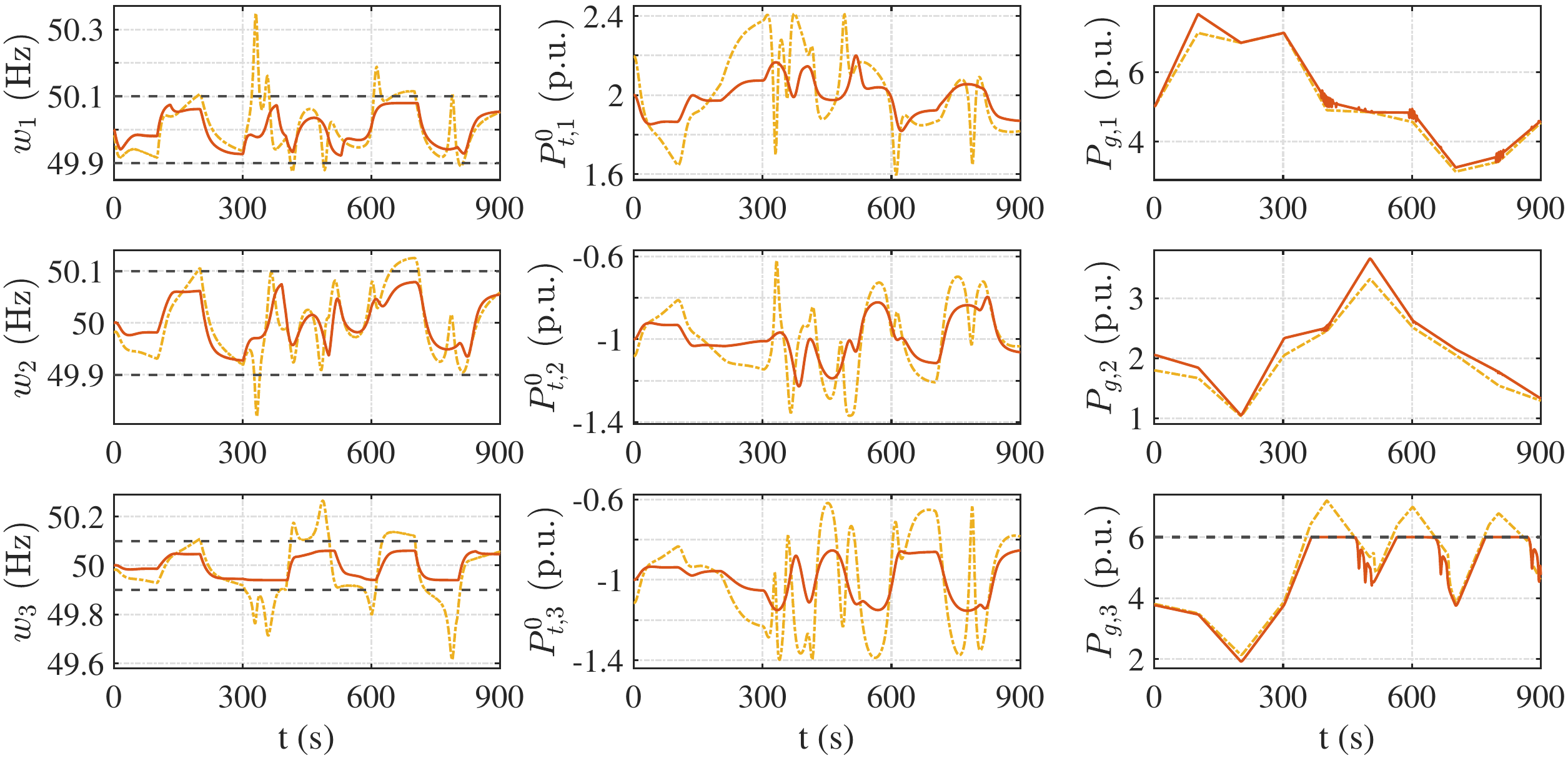}
  \caption{Control performance comparison under the time-varying net load, where yellow and red lines denote the responses under SFC and the proposed method, respectively. Left column: frequency responses of areas 1, 2, and 3 (black dashed lines denote safe frequency bounds). Middle column: net tie-line powers of areas 1, 2, and 3. Right column: generation outputs of areas 1, 2, and 3 (black dashed line denotes the upper generation capacity limit).}
  \label{Fig:Compare_TimeVarying}
\end{figure*}

The simulation results are presented in Fig.~\ref{Fig:Compare_TimeVarying}. It shows that the proposed control method successfully restores system frequency and net tie-line powers while maintaining the transient frequency safety and the capacity constraint. This confirms its robustness against net load prediction errors and parameter uncertainties (i.e., varied damping coefficient $D_i$ and rotational inertia $M_i$). Moreover, compared with SFC, the proposed method achieves smaller deviations in both frequency and net tie-line powers, which demonstrates the superior control performance of the proposed method. 

\section{Conclusion}
This paper has proposed a distributed frequency control scheme for multi-area power systems that explicitly enforces transient frequency safety constraints and capacity constraints while achieving optimal steady-state economic performance. The scheme has combined a feedback optimization layer for economic dispatch with a safety corrector layer for enforcing frequency safety and capacity limits during the transient procedure. Theoretical analysis has established existence and uniqueness of solutions, optimality of the steady state, asymptotic stability, and guaranteed transient frequency safety. Simulations on a 3-area test system have confirmed that, compared with conventional FO-based control, the proposed scheme has not only maintained system frequency within the safe region, but has also reduced deviations and accelerated the restoration of both system frequency and net tie-line powers to their nominal values.

\bibliographystyle{unsrt}        
\bibliography{SafeFO20250812.bib}           

\begin{thebibliography}{10}

\bibitem{7544628}
Y.~Ji, T.~Zheng, and L.~Tong.
\newblock Stochastic interchange scheduling in the real-time electricity
  market.
\newblock {\em IEEE Trans. Power Syst.}, 32(3):2017--2027, 2017.

\bibitem{9495219}
J.~Zhu, X.~Mo, Y.~Xia, Y.~Guo, J.~Chen, and M.~Liu.
\newblock Fully-decentralized optimal power flow of multi-area power systems
  based on parallel dual dynamic programming.
\newblock {\em IEEE Trans. Power Syst.}, 37(2):927--941, 2022.

\bibitem{kundur1994}
P.~Kundur.
\newblock {\em Power System Stability and Control}.
\newblock McGraw-Hill, New York, NY, USA, 1994.

\bibitem{Wood2014}
A.~J Wood, B.~F. Wollenberg, and G.~B. Shebl{\'e}.
\newblock {\em Power Generation, Operation, and Control}.
\newblock Wiley, New York, NY, USA, 3 rd edition, 2014.

\bibitem{DORFLER2017296}
F.~Dörfler and S.~Grammatico.
\newblock Gather-and-broadcast frequency control in power systems.
\newblock {\em Automatica}, 79:296--305, 2017.

\bibitem{Yan2020}
Z.~Yan and Y.~Xu.
\newblock A multi-agent deep reinforcement learning method for cooperative load
  frequency control of a multi-area power system.
\newblock {\em IEEE Trans. Power Syst.}, 35(6):4599--4608, 2020.

\bibitem{Zhao2022}
Y.~Zhao, T.~Liu, and D.~J. Hill.
\newblock A multi-agent reinforcement learning based frequency control method
  with data-enabled predictive control guided policy search.
\newblock In {\em Proc. IEEE PES General Meeting}, pages 1--5, Denver, CO, USA,
  2022.

\bibitem{10696981}
{Y. Zhao, T. Liu, and D. J. Hill}.
\newblock Distributed attention-enabled multi-agent reinforcement learning
  based frequency regulation of power systems.
\newblock {\em IEEE Trans. Power Syst.}, 40(3):2427--2437, 2025.

\bibitem{Yang2021b}
L.~Yang, T.~Liu, and D.~J. Hill.
\newblock Distributed {MPC}-based frequency control for multi-area power
  systems with energy storage.
\newblock {\em Electr. Power Syst. Res.}, 190:106642, 2021.

\bibitem{Liu2023}
X.~Liu, C.~Wang, X.~Kong, Y.~Zhang, W.~Wang, and K.~Y. Lee.
\newblock Tube-based distributed {MPC} for load frequency control of power
  system with high wind power penetration.
\newblock {\em IEEE Trans. Power Syst.}, 39(2):3118--3129, 2024.

\bibitem{Trip2016}
S.~Trip, M.~Bürger, and C.~{De Persis}.
\newblock An internal model approach to (optimal) frequency regulation in power
  grids with time-varying voltages.
\newblock {\em Automatica}, 64:240--253, 2016.

\bibitem{Trip2018}
S.~Trip and C.~De~Persis.
\newblock Distributed optimal load frequency control with non-passive dynamics.
\newblock {\em IEEE Trans. Control. Netw. Syst.}, 5(3):1232--1244, 2018.

\bibitem{9869334}
Y.~Jiang, W.~Cui, B.~Zhang, and J.~Cort{\'e}s.
\newblock Stable reinforcement learning for optimal frequency control: A
  distributed averaging-based integral approach.
\newblock {\em IEEE Open J. of Control Syst.}, 1:194--209, 2022.

\bibitem{Li2016}
N.~Li, C.~Zhao, and L.~Chen.
\newblock Connecting automatic generation control and economic dispatch from an
  optimization view.
\newblock {\em IEEE Trans. Control Netw. Syst.}, 3(3):254--264, 2016.

\bibitem{7944568}
E.~Mallada, C.~Zhao, and S.~Low.
\newblock Optimal load-side control for frequency regulation in smart grids.
\newblock {\em IEEE Trans. Autom. Control}, 62(12):6294--6309, 2017.

\bibitem{Ste2017}
T.~Stegink, C.~De~Persis, and A.~van~der Schaft.
\newblock A unifying energy-based approach to stability of power grids with
  market dynamics.
\newblock {\em IEEE Trans. Autom. Control}, 62(6):2612--2622, 2017.

\bibitem{Yang2021}
L.~Yang, T.~Liu, Z.~Tang, and D.~J. Hill.
\newblock Distributed optimal generation and load-side control for frequency
  regulation in power systems.
\newblock {\em IEEE Trans. Autom. Control}, 66(6):2724--2731, 2021.

\bibitem{10319778}
M.~Li, J.~Watson, and I.~Lestas.
\newblock Distributed optimal secondary frequency control in power networks
  with delay independent stability.
\newblock {\em IEEE Trans. Autom. Control}, 69(6):3748--3763, 2024.

\bibitem{zhao2024distributed}
X.~Zhao, Z.~Ma, S.~Zou, and X.~Shi.
\newblock Distributed optimal load frequency control for multi-area power
  systems with controllable loads.
\newblock {\em J. Franklin Inst.}, 361(13):107007, 2024.

\bibitem{wang2020distributed}
Z.~Wang, F.g Liu, C.~Zhao, Z.~Ma, and W.~Wei.
\newblock Distributed optimal load frequency control considering nonsmooth cost
  functions.
\newblock {\em Syst. \& Control Lett.}, 136:104607, 2020.

\bibitem{10488734}
Y.~Wang, S.~Liu, X.~Cao, and M.-Y. Chow.
\newblock An operator splitting scheme for distributed optimal load-side
  frequency control with nonsmooth cost functions.
\newblock {\em IEEE Trans. Autom. Control}, 69(9):6442--6449, 2024.

\bibitem{10085973}
Z.~Wang, W.~Wei, J.~Z.~F. Pang, F.~Liu, B.~Yang, X.~Guan, and S.~Mei.
\newblock Online optimization in power systems with high penetration of
  renewable generation: Advances and prospects.
\newblock {\em IEEE/CAA J. Autom. Sin.}, 10(4):839--858, 2023.

\bibitem{dorfler2023control}
F.~D{\"o}rfler and D.~Gro{\ss}.
\newblock Control of low-inertia power systems.
\newblock {\em Annu. Rev. Control Robot. Auton. Syst.}, 6(1):415--445, 2023.

\bibitem{8450880}
F.~Milano, F.~Dörfler, G.~Hug, D.~J. Hill, and G.~Verbič.
\newblock Foundations and challenges of low-inertia systems (invited paper).
\newblock In {\em Proc. 20th Power Syst. Comput. Conf.}, pages 1--25, Dublin,
  Ireland, 2018.

\bibitem{10005839}
T.~Zhao, J.~Wang, and M.~Yue.
\newblock A barrier-certificated reinforcement learning approach for enhancing
  power system transient stability.
\newblock {\em IEEE Trans. Power Syst.}, 38(6):5356--5366, 2023.

\bibitem{ZHANG2019274}
Y.~Zhang and J.~Cort{\'e}s.
\newblock Distributed transient frequency control for power networks with
  stability and performance guarantees.
\newblock {\em Automatica}, 105:274--285, 2019.

\bibitem{ZHANG2021109335}
{Y. Zhang and J. Cort{\'e}s}.
\newblock Model predictive control for transient frequency regulation of power
  networks.
\newblock {\em Automatica}, 123:109335, 2021.

\bibitem{9031310}
Y.~Zhang and J.~Cortés.
\newblock Distributed bilayered control for transient frequency safety and
  system stability in power grids.
\newblock {\em IEEE Trans. Control Netw. Syst.}, 7(3):1476--1488, 2020.

\bibitem{YUAN2024105753}
Z.~Yuan, C.~Zhao, and J.~Cortés.
\newblock Reinforcement learning for distributed transient frequency control
  with stability and safety guarantees.
\newblock {\em Syst. \& Control Lett.}, 185:105753, 2024.

\bibitem{7990543}
Z.~Wang, F.~Liu, S.~Low, C.~Zhao, and S.~Mei.
\newblock Distributed frequency control with operational constraints, part i:
  Per-node power balance.
\newblock {\em IEEE Trans. Smart Grid}, 10(1):40--52, 2019.

\bibitem{YI2016259}
P.~Yi, Y.~Hong, and F.~Liu.
\newblock Initialization-free distributed algorithms for optimal resource
  allocation with feasibility constraints and application to economic dispatch
  of power systems.
\newblock {\em Automatica}, 74:259--269, 2016.

\bibitem{9075378}
A.~Hauswirth, S.~Bolognani, G.~Hug, and F.~Dörfler.
\newblock Timescale separation in autonomous optimization.
\newblock {\em IEEE Trans. Autom. Control}, 66(2):611--624, 2021.

\bibitem{Cojocaru2004}
M.-G. Cojocaru and L.~Jonker.
\newblock Existence of solutions to projected differential equations in
  {Hilbert} spaces.
\newblock {\em Proc. Amer. Math. Soc.}, 132(1):183--193, 2004.

\bibitem{Hauswirth2021}
A.~Hauswirth, S.~Bolognani, and F.~D\"{o}rfler.
\newblock Projected dynamical systems on irregular, non-euclidean domains for
  nonlinear optimization.
\newblock {\em SIAM J. Control Optim.}, 59(1):635--668, 2021.

\bibitem{7937882}
P.~Glotfelter, J.~Cortés, and M.~Egerstedt.
\newblock Nonsmooth barrier functions with applications to multi-robot systems.
\newblock {\em IEEE Control Syst. Lett.}, 1(2):310--315, 2017.

\bibitem{boyd2004convex}
S.~Boyd and L.~Vandenberghe.
\newblock {\em Convex Optimization}.
\newblock Cambridge: Cambridge University Press, 2004.

\bibitem{FB-LNS}
F.~Bullo.
\newblock {\em Lectures on Network Systems}.
\newblock Kindle Direct Publishing, {1.7} edition, 2024.

\bibitem{Horn2013}
R.~A. Horn and C.~R. Johnson.
\newblock {\em Matrix Analysis}.
\newblock Cambridge Univ. Press, Cambridge, U.K., 2013.

\bibitem{8066371}
Y.~Wang, N.~Zhang, Q.~Chen, D.~S. Kirschen, P.~Li, and Q.~Xia.
\newblock Data-driven probabilistic net load forecasting with high penetration
  of behind-the-meter pv.
\newblock {\em IEEE Trans. Power Syst.}, 33(3):3255--3264, 2018.

\bibitem{1458194}
G.~Gross and F.~D. Galiana.
\newblock Short-term load forecasting.
\newblock {\em Proc. IEEE}, 75(12):1558--1573, 1987.

\bibitem{10454597}
C.~Si, H.~Wang, L.~Chen, J.~Zhao, Y.~Min, and F.~Xu.
\newblock Robust co-modeling for privacy-preserving short-term load forecasting
  with incongruent load data distributions.
\newblock {\em IEEE Trans. Smart Grid}, 15(3):2985--2999, 2024.

\bibitem{Khalil2002}
H.~K. Khalil.
\newblock {\em Nonlinear Systems}.
\newblock Prentice-Hall, Upper Saddle River, NJ, USA, 2002.

\bibitem{ENTSOE_FrequencyQuality_2015}
{ENTSO-E Nordic Analysis Group}.
\newblock Frequency quality - phase 1 report.
\newblock Technical report, 2015.

\bibitem{mosek}
{MOSEK ApS}.
\newblock {\em The MOSEK optimization toolbox for MATLAB manual. Version
  10.1.}, 2024.

\bibitem{6320655}
H.~Huang and F.~Li.
\newblock Sensitivity analysis of load-damping characteristic in power system
  frequency regulation.
\newblock {\em IEEE Trans. Power Syst.}, 28(2):1324--1335, 2013.

\end{thebibliography}

\end{document}